%% file: main.tex
\documentclass[final,1p,times]{elsarticle}

\input{Preamble/Preamble}
\begin{document}


\begin{frontmatter}

\input{Frontmatter/Title_authors_addresses/Title_authors_addresses}

\input{Frontmatter/Abstract_keyword/Abstract_keyword}

\end{frontmatter}


\input{Program_summary/Program_summary}

\section{Introduction}
\label{section: Introduction}
\input{Text/Introduction/Introduction}

\section{Basics of the program}
\label{section: Basics of the program}
\input{Text/Basics_of_the_program/Basics_of_the_program}

\subsection{Package of the program}
\label{subsection: Package of the program}
\input{Text/Basics_of_the_program/Package_of_the_program/Package_of_the_program}

\subsection{Input of the program}
\label{subsection: Input of the program}
\input{Text/Basics_of_the_program/Input_of_the_program/Input_of_the_program}

\subsection{Algorithm of the program}
\label{subsection: Algorithm of the program}
\input{Text/Basics_of_the_program/Algorithm_of_the_program/Algorithm_of_the_program}

\subsection{Execution of the program}
\label{subsection: Execution of the program}
\input{Text/Basics_of_the_program/Execution_of_the_program/Execution_of_the_program}

\subsection{Performance of the program}
\label{subsection: Performance of the program}
\input{Text/Basics_of_the_program/Performance_of_the_program/Performance_of_the_program}

\subsection{Output of the program}
\label{subsection: Output of the program}
\input{Text/Basics_of_the_program/Output_of_the_program/Output_of_the_program}

\subsection{Validation of the program}
\label{subsection: Validation of the program}
\input{Text/Basics_of_the_program/Validation_of_the_program/Validation_of_the_program}

\section{Component analysis}
\label{section: Component analysis}
\input{Text/Component_analysis/Component_analysis}

\subsection{Decay trees}
\label{subsection: Component analysis --- decay trees}
\input{Text/Component_analysis/Decay_trees/Decay_trees}

\subsection{Decay initial-final states}
\label{subsection: Component analysis --- decay initial-final states}
\input{Text/Component_analysis/Decay_initial-final_states/Decay_initial-final_states}

\subsection{Decay branches of particles}
\label{subsection: Component analysis --- decay branches of particles}
\input{Text/Component_analysis/Decay_branches_of_particles/Decay_branches_of_particles}

%

\subsection{Inclusive decay branches}
\label{subsection: Component analysis --- inclusive decay branches}
\input{Text/Component_analysis/Inclusive_decay_branches/Inclusive_decay_branches}


\subsection{Essential topology tags}
\label{subsection: Component analysis --- Essential topology tags}
\input{Text/Component_analysis/Essential_topology_tags/Essential_topology_tags}

\section{Signal identification}
\label{section: Signal identification}
\input{Text/Signal_identification/Signal_identification}

\subsection{Decay trees}
\label{subsection: Signal identification --- decay trees}
\input{Text/Signal_identification/Decay_trees/Decay_trees}

\subsection{Decay initial-final states}
\label{subsection: Signal identification --- decay initial-final states}
\input{Text/Signal_identification/Decay_initial-final_states/Decay_initial-final_states}

\subsection{Particles}
\label{subsection: Signal identification --- particles}
\input{Text/Signal_identification/Particles/Particles}

\subsection{Decay branches}
\label{subsection: Signal identification --- decay branches}
\input{Text/Signal_identification/Decay_branches/Decay_branches}




\subsection{Essential topology tags}
\label{subsection: Signal identification --- Essential topology tags}
\input{Text/Signal_identification/Essential_topology_tags/Essential_topology_tags}

\section{Common settings}
\label{section: Common settings}
\input{Text/Common_settings/Common_settings}

\subsection{Settings on input entries}
\label{subsection: Settings on input entries}
\input{Text/Common_settings/Settings_on_input_entries/Settings_on_input_entries}

\subsection{Setting on input decay branches}
\label{subsection: Setting on input decay branches}
\input{Text/Common_settings/Setting_on_input_decay_branches/Setting_on_input_decay_branches}



\subsection{Setting on charge conjugation}
\label{subsection: Setting on charge conjugation}
\input{Text/Common_settings/Setting_on_charge_conjugation/Setting_on_charge_conjugation}

\subsection{Setting on initial state particles}
\label{subsection: Setting on initial state particles}
\input{Text/Common_settings/Setting_on_initial_state_particles/Setting_on_initial_state_particles}

\clearpage

\section{Summary}
\label{section: Summary}
\input{Text/Summary/Summary}

\section*{Acknowledgements}
\input{Acknowledgements/Acknowledgements}

\section*{References}
\input{References/References}

\end{document}

%% file: Preamble/Preamble.tex
\usepackage{graphics}

\usepackage{amssymb}

\newcounter{bla}

\journal{Computer Physics Communications}

\usepackage{amsmath}
\usepackage{multirow}
\usepackage[colorlinks,linkcolor=blue]{hyperref}
\usepackage{subfigure}
\usepackage{longtable}
\usepackage{makecell}
\usepackage{color}
\newcommand{\tablecaption}[1]{\caption{#1} \\}
\usepackage{caption}
\newcommand{\tableheader}[1]
{
  \hline
  #1
  \hline
  \endfirsthead

  \hline
  #1
  \hline
  \endhead

  \endfoot

  \endlastfoot
}
\setcellgapes[t]{2pt}
\makegapedcells
\newcounter{rownumbers}
\newcommand\rn{\stepcounter{rownumbers}\arabic{rownumbers}}
\newcommand{\EOL}{\\}
\newcommand{\EOLTWO}{\\ \hline}
\newcommand{\topoTags}[1]{#1}

%% file: Frontmatter/Title_authors_addresses/Title_authors_addresses.tex
\title{TopoAna: A generic tool for the event type analysis of \\ inclusive Monte-Carlo samples in high energy \\ physics experiments}

\author[a]{Xingyu Zhou\corref{author}}
\author[b]{Shuxian Du}
\author[c]{Gang Li}
\author[d]{Chengping Shen\corref{author}}

\cortext[author] {Corresponding author. \\ \textit{E-mail address:} zhouxy@buaa.edu.cn, shencp@fudan.edu.cn}
\address[a]{School of Physics, Beihang University, Beijing 100191, China}
\address[b]{School of Physics and Microelectronics, Zhengzhou University, Zhengzhou 450000, China}
\address[c]{Institute of High Energy Physics, Chinese Academy of Sciences, Beijing 100049, China}
\address[d]{Key Laboratory of Nuclear Physics and Ion-beam Application (MOE) and Institute of Modern Physics, Fudan University, Shanghai 200443, China}

%% file: Frontmatter/Abstract_keyword/Abstract_keyword.tex
\begin{abstract}
Inclusive Monte-Carlo samples are indispensable for signal selection and background suppression in many high energy physics experiments.
A clear knowledge of the physics processes involved in the samples, including the types of processes and the number of processes in each type, is a great help to investigating signals and backgrounds.
To help analysts obtain the physics process information from the truth information of the samples, we develop a physics process analysis program, TopoAna, with C++, ROOT, and LaTeX.
The program implements the functionalities of component analysis and signal identification with many kinds of fine, customizable classification and matching algorithms.
It tags physics processes in individual events accurately in the output root files, and exports the physics process information at the sample level clearly to the output plain text, tex source, and pdf files.
Independent of specific software frameworks, the program is applicable to many experiments.
At present, it has come into use in three $e^+e^-$ colliding experiments: the BESIII, Belle, and Belle~II experiments.
The use of the program in other similar experiments is also prospective.
\end{abstract}

\begin{keyword}
event type; component analysis; signal identification; inclusive Monte-Carlo samples; high energy physics experiments
\end{keyword}

%% file: Program_summary/Program_summary.tex
\begin{small}

\noindent
{\bf PROGRAM SUMMARY}

\

\noindent
{\em Program title: TopoAna} \\
{\em Licensing provisions: MIT} \\
{\em Programming language: C++} \\
{\em Operating system: Linux} \\
{\em Nature of problem: A clear knowledge of the physics processes involved in inclusive Monte-Carlo samples is a great help to investigating signals and backgrounds in many high energy physics experiments. However, the raw topology truth information of the samples is counter-intuitive, diverse, and overwhelming, which makes it difficult for analysts to check the physics process information of the samples directly.} \\
{\em Solution method: Based on accurate pattern matching, many kinds of fine, customizable classification and matching algorithms are implemented in this program, in order to help analysts obtain the physics process information of the samples from their raw truth information.} \\
{\em Unusual features: Besides the C++ Standard Template Library, this program makes use of ROOT~[1], a C++ based data analysis software universally used in modern high energy physics experiments. In addition, the program employs the Linux command, pdflatex, to compile the tex source files into the pdf documents.} \\

\noindent \footnotesize{[1] ROOT User's Guide, Available online: \href{https://root.cern/root/htmldoc/guides/users-guide/ROOTUsersGuide.html}{https://root.cern/root/htmldoc/guides/users-guide/ROOTUsersGuide.html}.}

\end{small}

%% file: Text/Introduction/Introduction.tex
One of the most important tasks in the data analysis of high energy physics experiments is to select signals, or in other words, to suppress backgrounds.
As for the task, inclusive/generic Monte-Carlo (MC) samples are extremely useful, in that they provide basic, though not perfect, descriptions of the signals and/or backgrounds involved.
However, due to the similarities between signals and some backgrounds, it usually takes efforts to establish a set of selection criteria that retain a high signal efficiency and meanwhile keep a low background level. Further optimization of preliminary criteria is often needed in the process.
Under the circumstances, a comprehensive understanding of the samples is required.
In particular, a clear knowledge of the physics processes, or event types, involved in the samples is quite helpful.
To be specific, the physics process information includes the types of processes and the number of processes in each type, involved both in the entire samples and in the individual events.
Here, the physics process could be a complete production and decay process involved in an event, or merely a part of it, such as the decay of an intermediate resonance.
With the information, one can figure out the main backgrounds (especially the peaking ones), and optimize the selection criteria further by analyzing the differences between the main backgrounds and the signals.
Even if it is difficult to further suppress these backgrounds, the knowledge of their types is beneficial to estimate the systematic uncertainties associated with them.

The analysis of the physics process information described above is a sort of component analysis.
It is complex since it has to classify physics processes actively and finely.
Another sort of physics process analysis often required in practice is signal identification, which only aims to search for certain processes of interests.
It is relatively simple because its core technique is merely pattern matching.
Mostly, signal and background events coexist in inclusive MC samples.
It is useful to differentiate them in such cases.
The identified signal events can be used to make up a signal sample in the absence of specialized signal samples, or they can be removed to avoid repetition in the presence of specialized signal samples. Occasionally, we have to pick out some decay branches in order to re-weight them according to new theoretical predictions or updated experimental measurements. Signal identification also plays a part in this occasion.

\begin{figure}[htbp!]
\centering
\includegraphics[width=\textwidth]{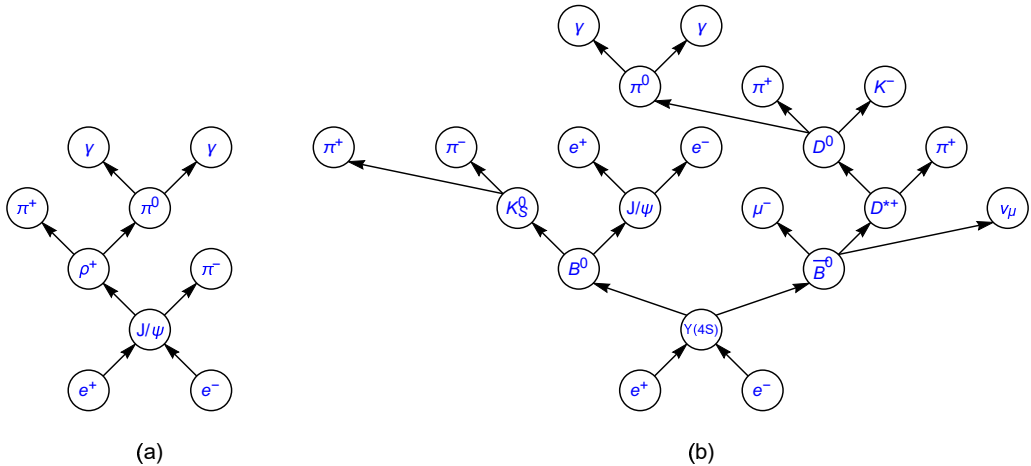}
\caption{Topology diagrams of (a) $e^+ e^- \to J/\psi$, $J/\psi \to \rho^+ \pi^-$, $\rho^+ \to \pi^+ \pi^0$, $\pi^0 \to \gamma \gamma$ and (b) $e^+ e^- \to \Upsilon(4S)$, $\Upsilon(4S) \to B^0 \bar{B}^0$, $B^0 \to K_S^0 J/\psi$, $\bar{B}^0 \to \mu^- D^{*+} \nu_{\mu}$, $K_S^0 \to \pi^+ \pi^-$, $J/\psi \to e^+ e^-$, $D^{*+} \to D^0 \pi^+$, $D^0 \to \pi^0 \pi^+ K^-$, $\pi^0 \to \gamma \gamma$. As if trees grow, the diagrams are plotted from bottom to top.}
\label{figure: Topology diagrams}
\end{figure}

Processes in high energy physics can be visualized with topology diagrams.
As an example, Fig.~\ref{figure: Topology diagrams} shows the topology diagrams of two typical physics processes occurring at $e^+e^-$ colliders.
From the figure, the hierarchies of the processes and the relationships among the particles are clearly illustrated with the diagrams.
Though the complexities of topology diagrams vary with physics processes, there is only one diagram corresponding to each process.
For this reason, we refer to the physics process information/analysis mentioned thereinbefore as topology information/analysis hereinafter. The component analysis and signal identification introduced above are exactly the two categories of topology analysis that will be discussed in this paper.

Since the raw topology truth information of inclusive MC samples is counter-intuitive, diverse, and overwhelming, it is difficult for analysts to check the topology information of the samples directly.
To help them do the checks quickly and easily, a topology analysis program called TopoAna is developed with C++, ROOT~\cite{ROOT1}, and LaTeX.
Here, C++ is the programming language, ROOT is the C++ based data analysis software universally used in modern high energy physics experiments, and LaTeX is used for generating pdf documents containing the obtained topology information.
The program implements the functionalities of component analysis and signal identification based on accurate pattern matching.
To meet a variety of practical requirements, many kinds of fine, customizable classification and matching algorithms are implemented in the program.
Generally, the program recognizes, categorizes, and counts physics processes in each event in the samples, and tags them in the corresponding entry of the output root (TFile~\cite{ROOT2}) files.
After processing the events, the program exports the obtained topology information at the sample level to the output plain text, tex source, and pdf files.

The program is applicable to inclusive MC samples at any data analysis stage of associated high energy physics experiments.
In the overwhelming majority of situations, it is run over the samples which have undergone some selections, in order to examine the signals and backgrounds in the selected samples as well as the effect of the imposed selections.
In such situations, the results of topology analysis are usually used together with other quantities for physics analysis.
In spite of this, applying the program to the samples without undergoing any selection facilitates us to validate the generators and decay cards that produce the samples and helps novices get familiar with the topology information of the samples.

The program has a history of more than ten years.
It has already gone through a series of major upgrades.
Prior to its development, analysts usually wrote some private codes to match few signals and/or backgrounds for their own studies.
The limited functions of these codes do not satisfy the increasing demand for topology analysis.
This motivates us to develop a generic, powerful, and easy-to-use program.
At first, the program was developed for the BESIII experiment, an experiment in the $\tau$-Charm energy region with abundant research topics under study~\cite{BESIII1, BESIII2}.
Later, it was extended substantially for the Belle~II experiment, which is primarily dedicated to search for physics beyond the Standard Model in the flavor sector and has already started data taking in the recent three years~\cite{Belle II}.
Besides, the program has also been tried and used in the Belle experiment, the predecessor of the Belle~II experiment, where some physics studies are still ongoing~\cite{Belle}.
Not relying on any specific software frameworks, the program now applies to many high energy physics experiments.

This paper gives an essential description of TopoAna.
It proceeds as follows:
Section \ref{section: Basics of the program} introduces the basics of the program;
Sections \ref{section: Component analysis} and \ref{section: Signal identification} expatiate the two categories of functionalities of the program --- component analysis and signal identification, respectively;
Section \ref{section: Common settings} presents some common settings for the executing of the program;
Section \ref{section: Summary} summarizes the paper.
It is worth mentioning here that, aside from the essential description in the paper, a detailed description of the program can be found in the file ``user\_guide\_v*.pdf'' under the directory ``share'' of the package.

%% file: Text/Basics_of_the_program/Basics_of_the_program.tex
This section introduces the basics of the program, including the package, input, algorithm, execution, performance, output, and validation of the program.
The package implements the program via a C++ class called ``topoana'' and a main function invoking the class.
Compiling the package creates the executable file of the program, that is, ``topoana.exe''.
To execute the program, we have to first obtain the input data of the program, namely the raw topology truth information of the inclusive MC samples, with some interfaces to the program in the software systems of the corresponding experiments.
Normally, the input data contain all the topology information of the samples.
With the data, all kinds of the topology analysis presented in the paper can be performed.

To carry out the topology analysis desired in our work, we have to provide some necessary input, functionality, and output information to the program.
The information is required to be filled in the setting items designed and implemented in the program, and the items have to be put in a plain text file named with a suffix ``.card''.
With the card file, one can execute the program with the command line: ``topoana.exe cardFileName'', where the argument ``cardFileName'' is optional and its default value is ``topoana.card''.
After the execution of the program, we can examine the results of topology analysis in the output files and use them to analyze other experimental quantities.
The results help us gain a better understanding of the signals and backgrounds and are conducive to carrying our work forward.
Besides the package, input, execution, and output of the program mentioned above, the algorithm, performance, and validation of the program will also be discussed in this section, because they are also essential aspects of the program.
In the next seven subsections, we will present the package, input, algorithm, execution, performance, output, and validation of the program in detail, with each part in one subsection.

%% file: Text/Basics_of_the_program/Package_of_the_program/Package_of_the_program.tex
The package consists of six directories --- ``include'', ``src'', ``bin'', ``share'', ``examples'', and ``utilities'' --- and five files --- ``LICENSE'', ``README.md'', ``Configure'', ``Makefile'', and ``Setup''.
While the directory ``include'' only includes one header file ``topoana.h'', the directory ``src'' contains sixty source files ``*.cpp'' as well as a script file ``topoana.C''.
At present, only one class, namely ``topoana'', is defined in the program for all of its functionalities.
The class is declared in ``topoana.h'', implemented in ``*.cpp'' files, and invoked in ``topoana.C''.

The file ``template\_topoana.card'' under the directory ``share'' saves all the items which are developed for users to specify information for the execution of the program.
One can refer to the file when filling in the cards for their own needs.
Some plain text files ``pid\_3pchrg\_txtpnm\_texpnm \_iccp.dat\_*'' are also included in the directory ``share''.
They store the basic information of the particles used in the program.
The suffixes of their names indicate the experiments they apply to.
One of them will be copied to ``pid\_3pchrg\_txtpnm\_texpnm\_iccp.dat'' when we set up the program.
Besides, the directory ``share'' also contains three LaTeX style files `` geometry.sty'', ``ifxetex.sty'', and ``makecell.sty'', which are invoked by the program for generating pdf files.
The directory ``examples'' includes plenty of detailed examples.
Particularly, all the examples involved in this paper are under its sub-directory ``in\_the\_paper''.
The directory ``utilities'' contains some useful bash scripts.

The program is released under MIT license~\cite{MIT license}.
The file ``README.md'' briefly introduces how to install and use the program.
To set up the program, one should first set the package path with the command ``./Configure''.
Standard outputs of the command are the guidelines for manually adding the absolute path of ``topoana.exe'' to the environment variable ``PATH'', in order to execute it without any path.
The second step is executing the command ``make''.
This command compiles the header, source, and script files into the executable file ``topoana.exe'' under the directory ``bin'', according to the rules specified in the ``Makefile''.
The last step is specifying the experiment name with the command line ``./Setup experimentName''.
Currently, the supported experiment names are ``BESIII'', ``Belle'', and ``Belle\_II''.
Besides, ``./Setup Example'' is required for the execution of the examples in the paper.

%% file: Text/Basics_of_the_program/Input_of_the_program/Input_of_the_program.tex
The input of the program is one or more root files including a TTree~\cite{ROOT2} object which has some TBranch~\cite{ROOT2} objects containing the raw topology truth information of the inclusive MC samples under study.
To be specific, the information in each entry of the TTree object consists of the following three ingredients associated with the particles produced in an event of the samples: the number of particles, PDG~\cite{PDG} codes of particles, and mother indices of particles.
Notably, the particles do not include the initial state particles ($e^+$ and $e^-$ in $e^+e^-$ colliding experiments), which are default and thus omitted.
Besides, the indices of particles are integers starting from zero (included) to the number of particles (excluded); they are obvious and hence not taken as an input ingredient for topology analysis.
Equation~(\ref{equation: Raw topology data of a specific MC event}) shows an example of the input data.

\begin{equation}
\footnotesize
\begin{array}{llll}
 & \text{Number of particles}           & : & \text{63} \\
 & \text{PDG codes of particles}        & : & \text{300553,} \\
 &                        &   & \text{$-$511, 511, $-$433, 421, 211, 22, $-$413, 111, 111, 113,} \\
 &                        &   & \text{211, $-$431, 22, $-$323, 213, $-$421, $-$211, 22, 22, 22,} \\
 &                        &   & \text{22, 211, $-$211, 333, 11, $-$12, 22, $-$311, $-$211, 211,} \\
 &                        &   & \text{111, 221, 331, 321, $-$321, 310, 22, 22, 111, 111,} \\
 &                        &   & \text{111, 111, 111, 221, 111, 111, 22, 22, 22, 22,} \\
 &                        &   & \text{22, 22, 22, 22, 22, 22, 22, 22, 22, 22,} \\
 &                        &   & \text{22, 22} \\
 & \text{Mother indices of particles}  & : & \text{$-$1,} \\
 &                        &   & \text{0, 0, 1, 1, 1, 1, 2, 2, 2, 2,} \\
 &                        &   & \text{2, 3, 3, 4, 4, 7, 7, 8, 8, 9,} \\
 &                        &   & \text{9, 10, 10, 12, 12, 12, 12, 14, 14, 15,} \\
 &                        &   & \text{15, 16, 16, 24, 24, 28, 31, 31, 32, 32,} \\
 &                        &   & \text{32, 33, 33, 33, 36, 36, 39, 39, 40, 40,} \\
 &                        &   & \text{41, 41, 42, 42, 43, 43, 44, 44, 45, 45,} \\
 &                        &   & \text{46, 46}
\end{array}
\label{equation: Raw topology data of a specific MC event}
\end{equation}

\noindent The complete physics process contained in the data is displayed as follows.
\begin{equation}
\footnotesize
\begin{array}{lllrllllllr}
 & 0 & e^{+} e^{-} \rightarrow \Upsilon(4S) & $-$1 & & & & & 9 & \rho^{+} \rightarrow \pi^{0} \pi^{+} & 6 \\
 & 1 & \Upsilon(4S) \rightarrow B^{0} \bar{B}^{0} & 0 & & & & & 10 & K^{*-} \rightarrow \pi^{-} \bar{K}^{0} & 6 \\
 & 2 & B^{0} \rightarrow \pi^{0} \pi^{0} \rho^{0} \pi^{+} D^{*-} & 1 & & & & & 11 & D_{s}^{-} \rightarrow e^{-} \bar{\nu}_{e} \phi \gamma & 7 \\
 & 3 & \bar{B}^{0} \rightarrow \pi^{+} D^{0} D_{s}^{*-} \gamma & 1 & & & & & 12 & \eta \rightarrow \pi^{0} \pi^{0} \pi^{0} & 8 \\
 & 4 & \rho^{0} \rightarrow \pi^{+} \pi^{-} & 2 & & & & & 13 & \eta^{\prime} \rightarrow \pi^{0} \pi^{0} \eta & 8 \\
 & 5 & D^{*-} \rightarrow \pi^{-} \bar{D}^{0} & 2  & & & & & 14 & \bar{K}^{0} \rightarrow K_{S}^{0} & 10 \\
 & 6 & D^{0} \rightarrow \rho^{+} K^{*-} & 3 & & & & & 15 & \phi \rightarrow K^{+} K^{-} & 11 \\
 & 7 & D_{s}^{*-} \rightarrow D_{s}^{-} \gamma & 3 & & & & & 16 & \eta \rightarrow \gamma \gamma & 13 \\
 & 8 & \bar{D}^{0} \rightarrow \eta \eta^{\prime} & 5 & & & & & 17 & K_{S}^{0} \rightarrow \pi^{0} \pi^{0} & 14
\end{array}
\label{equation: Decay tree of a specific MC event}
\end{equation}

\noindent Here, the decay branches in the process are placed into two blocks in order to make full use of the page space.
In both blocks, the first, second, and third columns are the indices, symbolic expressions, and mother indices of the decay branches.
Notably, all the decay branches of $\pi^0 \to \gamma \gamma$ are omitted in Eq.~(\ref{equation: Decay tree of a specific MC event}) in order to make the process look more concise.
Since the topology diagram of such a process looks like a tree, we refer to the complete processes as decay trees.
Obviously, the input data do not show the structure automatically.
Thus, we need the program to do the topology analysis work.

From the first branch in Eq.~(\ref{equation: Decay tree of a specific MC event}), only one particle $\Upsilon(4S)$ is produced after the $e^+e^-$ annihilation.
Thus, $\Upsilon(4S)$ can be referred to as the root particle of the decay tree.
Similarly, many other resonances with the quantum numbers $J^{PC}=1^{--}$, such as $J/\psi$, can be solely produced at other proper energy points.
Besides the cases with only one root particle, the program can deal with the cases with multiple root particles.
For example, the program can recognize the following raw topology truth information
\begin{equation}
\footnotesize
\begin{array}{llll}
 & \text{Number of particles}           & : & \text{25} \\
 & \text{PDG codes of particles}        & : & \text{433,} \\
 &                        &   & \text{$-$321, 223, 211, $-$413, 431, 111, 211, $-$211, 111, $-$411,} \\
 &                        &   & \text{111, 321, 113, 22, 22, 22, 22, 321, $-$211, $-$211,} \\
 &                        &   & \text{22, 22, 211, $-$211} \\
 & \text{Mother indices of particles}  & : & \text{$-$1,} \\
 &                        &   & \text{$-$1, $-$1, $-$1, $-$1, 0, 0, 2, 2, 2, 4,} \\
 &                        &   & \text{4, 5, 5, 6, 6, 9, 9, 10, 10, 10,} \\
 &                        &   & \text{11, 11, 13, 13}
\end{array}
\label{equation: Raw topology data of another specific MC event}
\end{equation}
\noindent as the following process
\begin{equation}
\footnotesize
\begin{array}{lllrllllllr}
 & 0 & e^{+} e^{-} \rightarrow \pi^{+} \omega K^{-} D^{*-} D_{s}^{*+} & -1 & & & & & 4 & D^{-} \rightarrow \pi^{-} \pi^{-} K^{+} & 2 \\
 & 1 & \omega \rightarrow \pi^{0} \pi^{+} \pi^{-} & 0 & & & & & 5 & D_{s}^{+} \rightarrow \rho^{0} K^{+} & 3 \\
 & 2 & D^{*-} \rightarrow \pi^{0} D^{-} & 0 & & & & & 6 & \rho^{0} \rightarrow \pi^{+} \pi^{-} & 5 \\
 & 3 & D_{s}^{*+} \rightarrow \pi^{0} D_{s}^{+} & 0 & & & & &   &  &   \\
\end{array}
\label{equation: Decay tree of another specific MC event}
\end{equation}
\noindent Here, the particles $\pi^{+} \omega K^{-} D^{*-} D_{s}^{*+}$ in the first branch arise from hadronization processes, in which quark pairs produced from initial state particles turn into hadrons.
The processes with hadronization ignored have a tree structure and thus are easy to resolve.
On the other hand, some hadronization processes, particularly those in high energy regions, contain complicated loop structures that are difficult to resolve without sophisticated algorithms.
Resolving these intricate hadronization processes is not involved in the program at present.

The input data are recommended to be saved in the TTree object together with other quantities for physics analysis, in order to facilitate the examination of the distributions of these quantities with the topology information.
It is easy to get the input of the program within the software framework of high energy physics experiments.
To facilitate its use, we have developed the interfaces of the program to the software systems of the BESIII, Belle, and Belle~II experiments.
Similar interfaces for other experiments can also be implemented with ease.
Beyond the scope of the paper, we will not discuss the details of the interfaces here.

%% file: Text/Basics_of_the_program/Algorithm_of_the_program/Algorithm_of_the_program.tex
The program resolves physics processes from the input data introduced above.
Considering the diversity of the data, the program first sorts them before translating them into physics processes.
Here, the diversity means that the data representing a process may have multiple permutations.
For example, the data for the decay $\rho^{0} \rightarrow \pi^{+} \pi^{-}$ have the following two permutations.
\begin{equation*}
\footnotesize
\begin{array}{lllllll}
 & \text{Number of particles}           & : & \text{3} \\
 & \text{PDG codes of particles}        & : & \text{\underline{113, 211, $-$211} or \underline{113, $-$211, 211}} \\
 & \text{Mother indices of particles}   & : & \text{$-$1, 0, 0}
\end{array}
\label{equation: Raw topology data of rho0 to pip and pim}
\end{equation*}
A decay tree can consist of many decay branches.
As a consequence, the diversity issue is complex.
To avoid the different permutations of one group of data are identified as different processes, the program first sorts the input data to adjust all the possible permutations to a unique order, according to the PDG codes and electronic charges of the involved particles, and the numbers of their daughter particles in the case of identical particles present in the same decay branch.
For example, the two permutations above will be finally sorted into the first permutation (113, 211, $-$211) in the program.
The sorting algorithm is implemented in the source file ``sortPs.cpp'', where some other settings are also involved.
One can see the reference file ``sortPs.cpp\_core'' for the core of the sorting algorithm.
After the sorting, the program can get the decay tree from the sorted data into a vector of the type ``vector\textless\ list\textless int\textgreater\ \textgreater'' with the function implemented in the source file ``getDcyTr.cpp''.

\begin{figure}[htbp!]
\centering
\includegraphics[width=\textwidth]{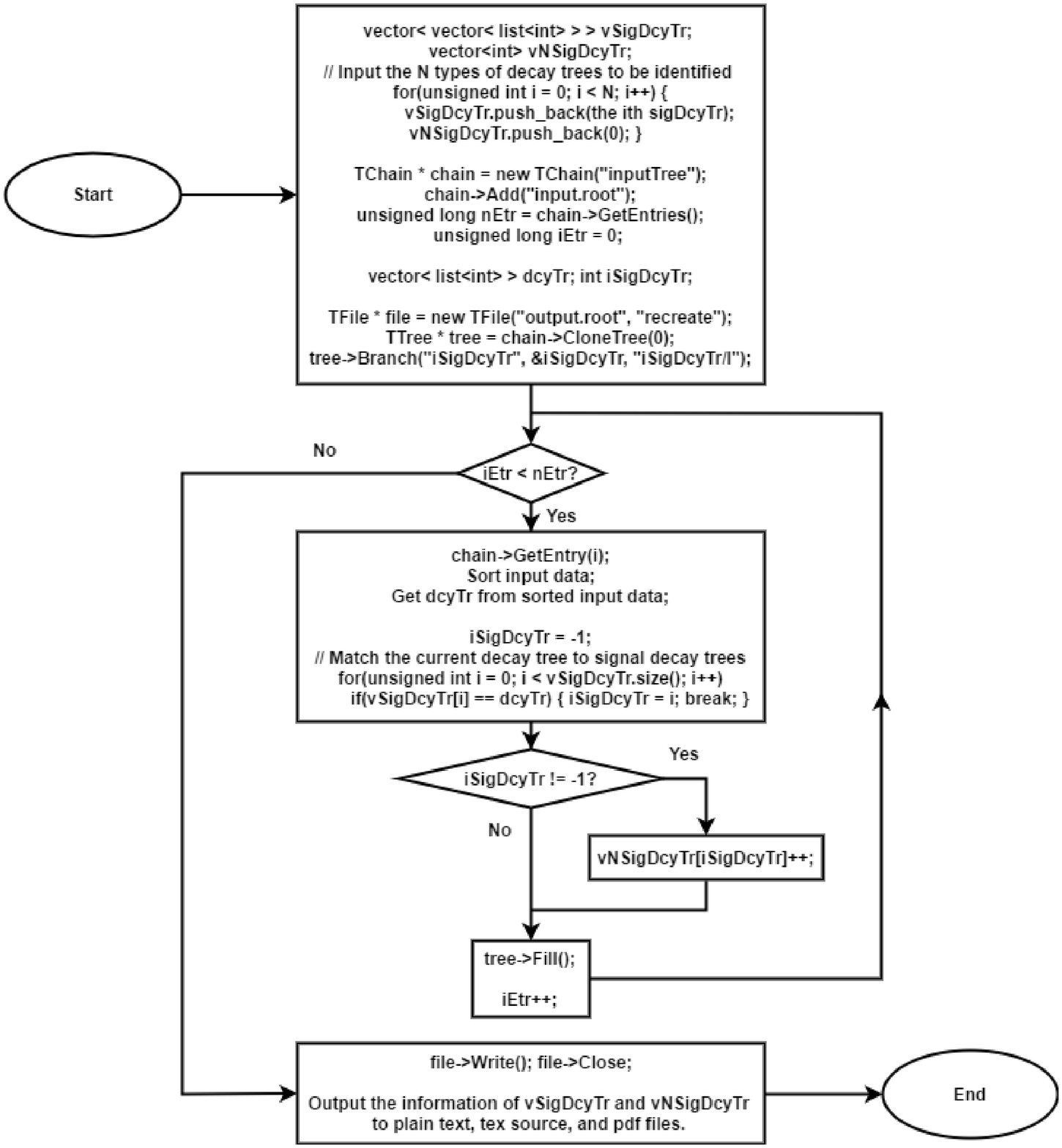}
\caption{Basic flow chart of the signal identification for decay trees.
The vectors ``vSigDcyTr'' and ``vNSigDcyTr'' are used to store the signal decay trees specified in the user card file and the numbers of these decay trees found in the input root file, respectively.
The TBranch ``iSigDcyTr'' in the output root file is used to record the index of the signal decay tree involved in each entry of the input root file.}
\label{figure: Basic flow chart of the signal identification for decay trees}
\end{figure}

\begin{figure}[htbp!]
\centering
\includegraphics[width=\textwidth]{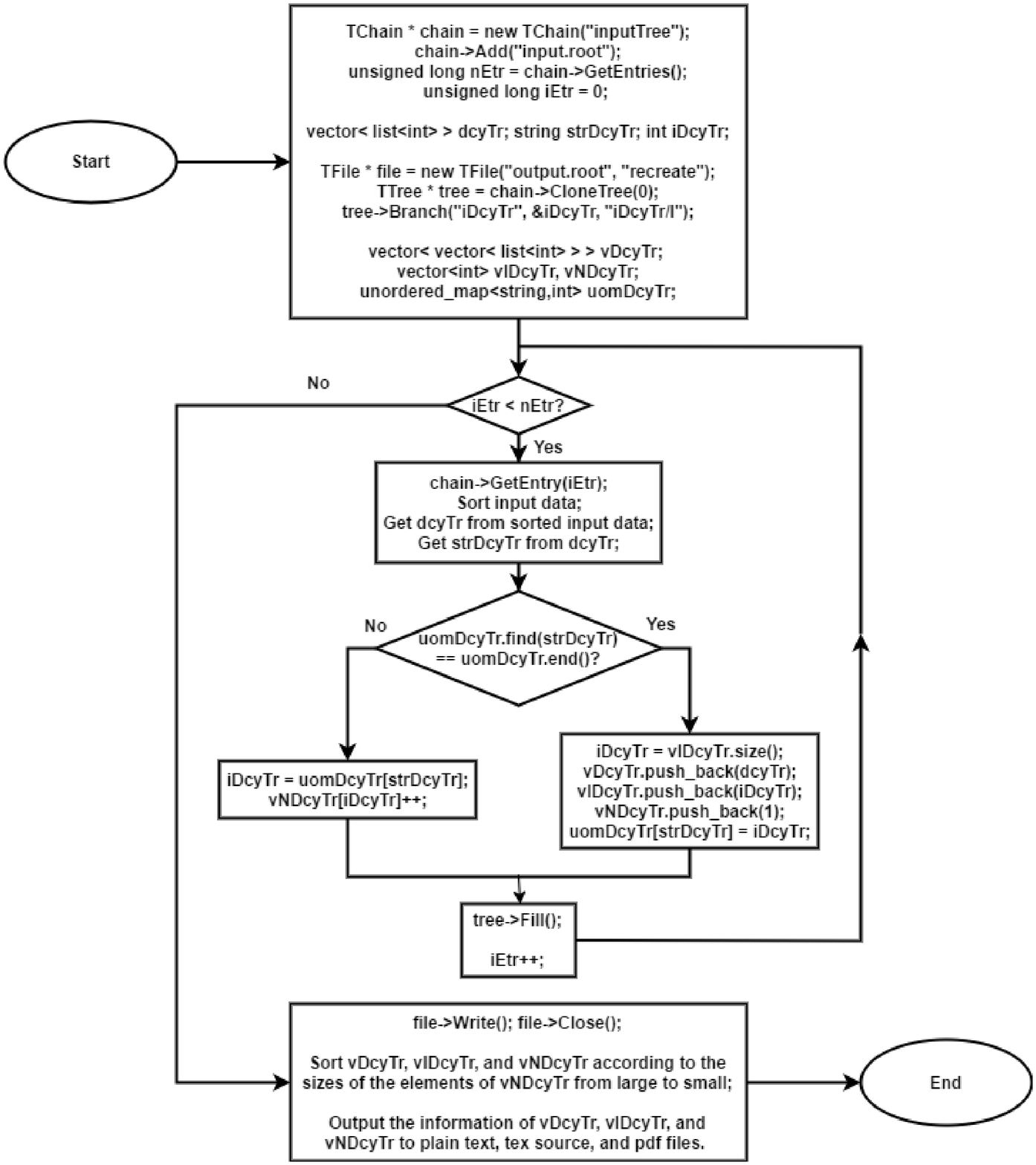}
\caption{Basic flow chart of the component analysis over decay trees.
The TBranch ``iDcyTr'' in the output root file is used to record the index of the decay tree involved in each entry of the input root file.
The vectors ``vDcyTr'', ``vIDcyTr'', and ``vNDcyTr'' are used to store the decay trees found in the input root file, their individual indices, and their individual numbers, respectively.
In addition, the unordered\_map ``uomDcyTr'' is used for the fast matching of decay trees.
Its key and value are the string ``strDcyTr'' and the index ``iDcyTr'', respectively.
Here, the string ``strDcyTr'' is constructed from the vector ``dcyTr''; there is a one-to-one correspondence between them.
}
\label{figure: Basic flow chart of the component analysis over decay trees}
\end{figure}

As mentioned in the previous section, the program has two categories of functionalities: signal identification and component analysis.
In this subsection, we introduce the basic algorithms for signal identification and component analysis by taking the cases of decay trees as examples.
Figures~\ref{figure: Basic flow chart of the signal identification for decay trees} and \ref{figure: Basic flow chart of the component analysis over decay trees} show the flow charts of these algorithms in detail.
Dozens of lines of code, including some using the ROOT classes TChain, TFile, and TTree~\cite{ROOT2}, are involved in the charts in order to express the algorithms explicitly.
The flow chart of the signal identification for decay trees is depicted in Fig.~\ref{figure: Basic flow chart of the signal identification for decay trees}.
Firstly, the program reads in the signal decay trees specified in the user card file.
Then, for each entry of the input root file, the program obtains the decay tree from the sorted input data, matches the decay tree to the signal decay trees, records the index of the matched signal decay tree, and increases the number of the matched signal decay tree.
At last, the program outputs the statistics of the signal decay trees.

The flow chart of the component analysis over decay trees is illustrated in Fig.~\ref{figure: Basic flow chart of the component analysis over decay trees}.
Despite the similarity in their frameworks, the flow chart has significant differences from that of the signal identification for decay trees in Fig.~\ref{figure: Basic flow chart of the signal identification for decay trees}.
In the signal identification algorithm, the signal decay trees to be identified are specified beforehand in the user card file.
On the contrary, in the component analysis algorithm, the program has to classify decay trees by itself from scratch.
In the signal identification algorithm, the decay trees are matched by directly comparing the vectors storing them.
Since the number of specified signal decay trees is fixed and usually small, the processing rate of the program is high and usually in constant.
However, in the component analysis algorithm, the number of decay tree types found in a sample can be quite large and tends to grow with the number of processed entries.
On this occasion, if we still match the decay trees by comparing the vectors storing them, the processing rate of the program will decrease with the increase of the number of processed entries.
To improve the processing rate, the unordered map~\cite{Unordered maps}, a kind of container template introduced since the C++ 11 standard, is employed for the fast matching of decay trees.
Internally, the elements in the unordered maps are organized into buckets depending on their hash values, to allow for fast access to individual elements directly by their key values with a constant average time complexity~\cite{Unordered maps}.
This constant feature in average time complexity will be examined in Section~\ref{subsection: Performance of the program}.

%% file: Text/Basics_of_the_program/Execution_of_the_program/Execution_of_the_program.tex
To execute the program, we have to first configure some necessary setting items in a card file, and then
 run the program with the command line: ``topoana.exe cardFileName''.
This subsection introduces the essential items for the input, basic functionality, and output of the program.
More items that can be set in the card file will be described in the following three sections.
Sections \ref{section: Component analysis} and \ref{section: Signal identification} expatiate the available items for the functionalities of the program, and Section \ref{section: Common settings} presents some optional items for the common settings to control the execution of the program.

\

An example of the card file containing the essential items is shown as follows.
\ \\
\begin{footnotesize}

\# The following five items set the input of the program.

\

\% Names of input root files

\{

\quad ../input/jpsi\_1.root

\quad ../input/jpsi\_2.root

\}

\

\% TTree name

\{

\quad evt

\}

\

\% TBranch name of the number of particles (Default: nMCGen)

\{

\quad Nmcps

\}

\

\% TBranch name of the PDG codes of particles (Default: MCGenPDG)

\{

\quad Pid

\}

\

\% TBranch name of the mother indices of particles (Default: MCGenMothIndex)

\{

\quad Midx

\}

\

\# The following item sets the basic functionality of the program.

\

\% Component analysis --- decay trees

\{

\quad Y

\}

\

\# The following item sets the output of the program.

\

\% Common name of output files (Default: Name of the card file)

\{

\quad jpsi\_ta

\}

\end{footnotesize}
\ \\
In the card file, ``\#'', ``\%'', and the pair of ``\{'' and ``\}'', are used for commenting, prompting, and grouping, respectively.
The first five, sixth, and last items are set for the input, basic functionality, and output of the program, respectively.

The first item sets the names of the input root files.
The names ought to be input one per line without tailing characters, such as comma, semicolon, and period.
In the names, both the absolute and relative paths are allowed and wildcards ``[]?*'' are supported, just like those in the root file names input to the method Add() of the class TChain~\cite{ROOT2}.
The second item specifies the TTree name.
The following three items set the TBranch names of the three ingredients of the raw topology truth information.
Of the first five items, the former two are indispensable, whereas the latter three can be removed or left empty if the input values are identical to the default values indicated in their prompts.

The sixth item sets the basic functionality of the program, namely the component analysis over decay trees.
The item can be replaced or co-exist with other functionality items expatiated in Sections \ref{section: Component analysis} and \ref{section: Signal identification}.
Here, we note that at least one functionality item has to be specified explicitly in the card file, otherwise the program will terminate soon after its start because no topology analysis to be performed is set up.

The last item specifies the common name of the output files. Though in different formats, the files are denominated with the same name for the sake of uniformity. They will be introduced in detail in the next subsection. This item is also optional, with the name of the card file as its default input value. It is a good practice to first denominate the card file with the desired common name of the output files and then remove this item or leave it empty.

To provide a complete description, we list and explain all the essential items in the paragraphs above.
However, in practical uses, we suggest removing the optional items if the input values are identical to the default ones.
In this way, the contents of the card file will become much more concise, making the use of the program easier and quicker.
For example, unless otherwise stated, only the following two items are used to set the essential information in Sections \ref{section: Component analysis}, \ref{section: Signal identification}, and \ref{section: Common settings}.
\ \\
\begin{footnotesize}

\% Names of input root files

\{

\quad ../input/mixed\_1.root

\quad ../input/mixed\_2.root

\}

\

\% TTree name

\{

\quad evt

\}

\end{footnotesize}

%% file: Text/Basics_of_the_program/Performance_of_the_program/Performance_of_the_program.tex
\begin{figure}[htbp!]
\centering
\subfigure{\includegraphics[width=0.475\textwidth]{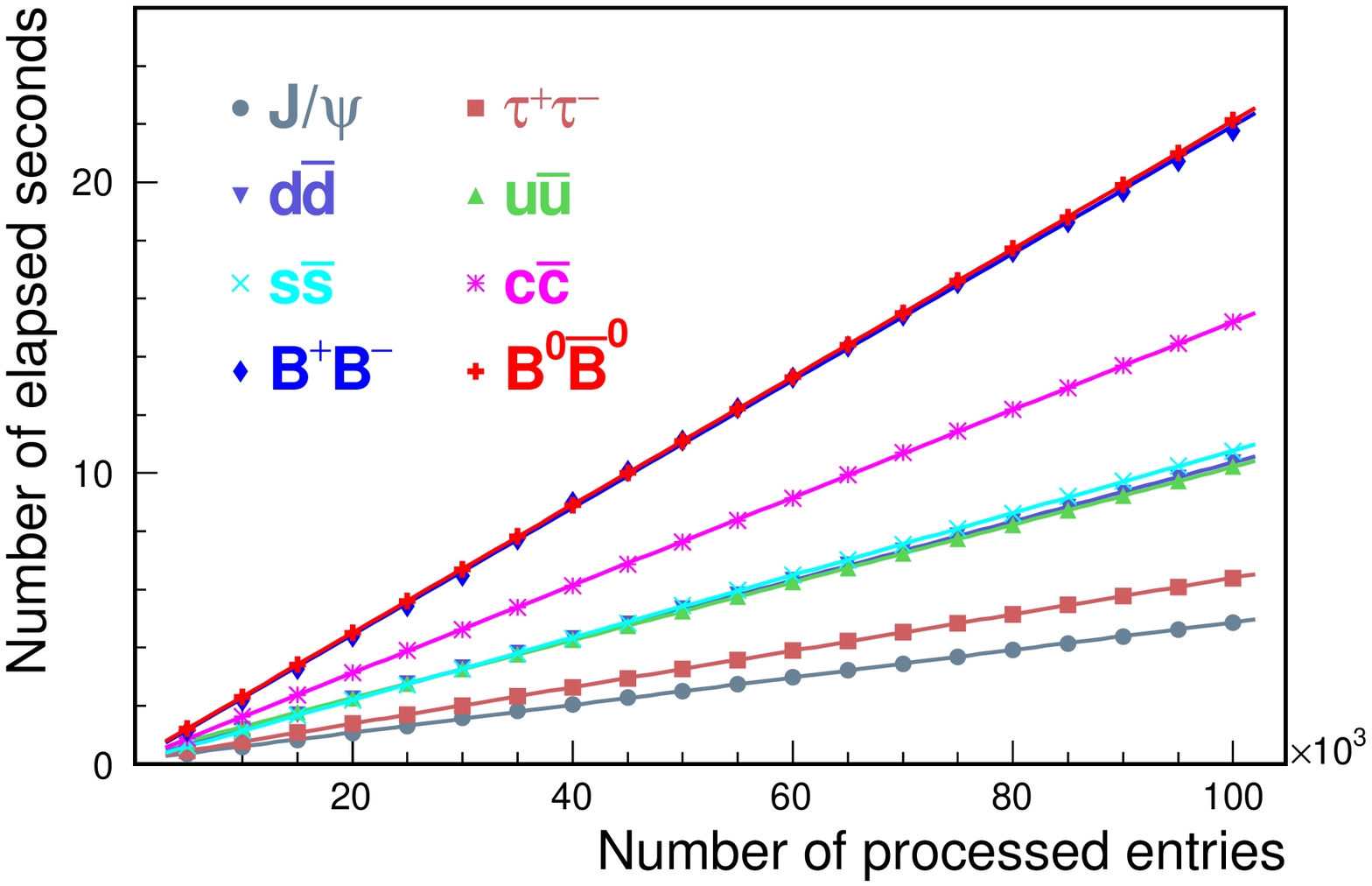}}
\subfigure{\includegraphics[width=0.475\textwidth]{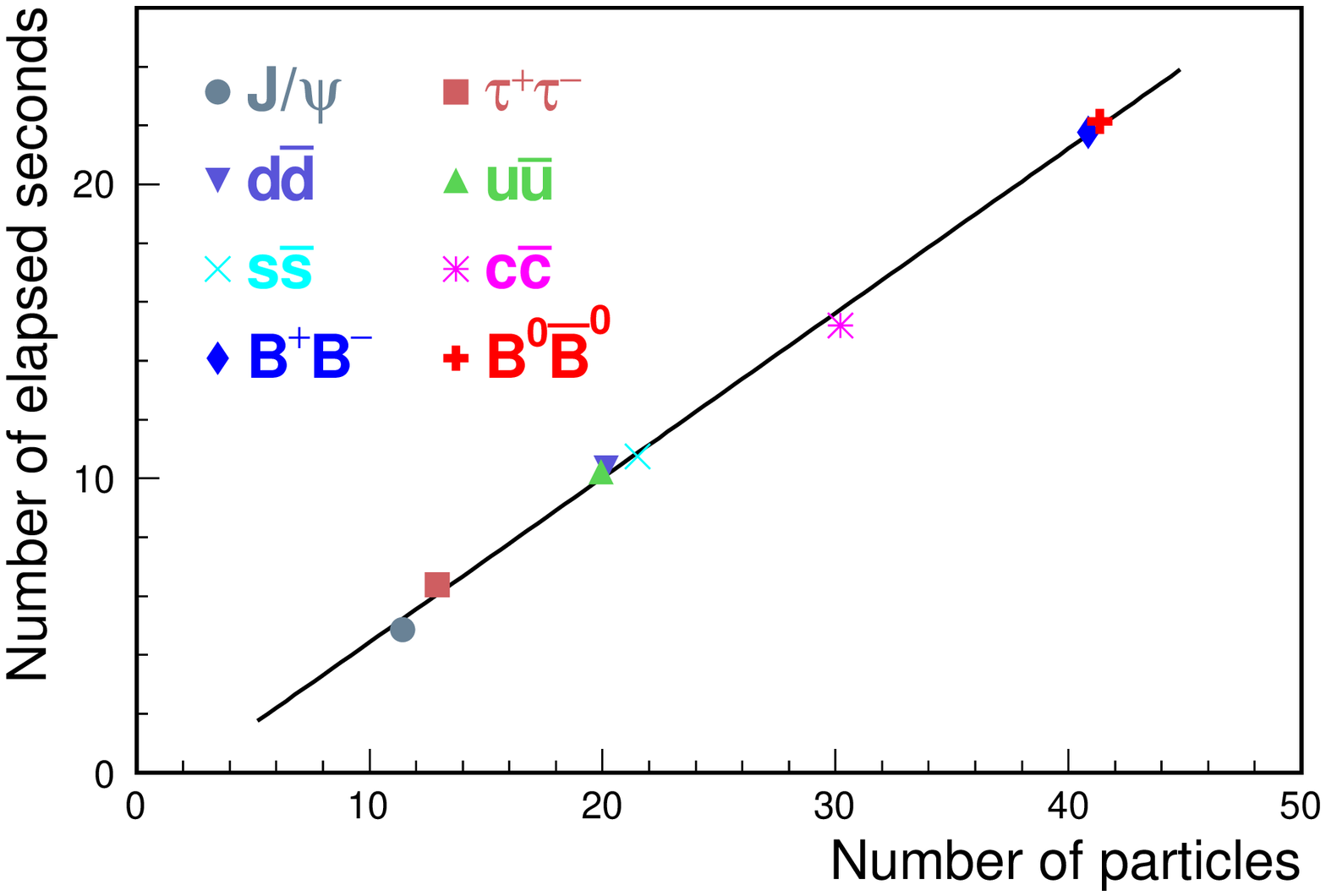}}
\caption{Performance study of the program with the $J/\psi$ sample as well as the $\tau^+\tau^-$, $d\bar{d}$, $u\bar{u}$, $s\bar{s}$, $c\bar{c}$, $B^+B^-$, and $B^0\bar{B}^0$ samples generated at the peak energy of the $\Upsilon(4S)$ resonance.
The left plot demonstrates the changing trends of the number of elapsed seconds with respect to the number of processed entries.
The right plot illustrates the relationship between the total number of elapsed seconds over the whole sample and the average number of generated particles in an event.
In both plots, the dots show the timing data from the standard output of the program, and the lines display the results of fitting linear functions to the data.
}
\label{figure: Performance study of the program}
\end{figure}

Besides the performance of the used computing systems, the processing rate of the program is largely related to the characteristics of the samples, particularly the average number of generated particles in each event.
Figure \ref{figure: Performance study of the program} shows the performance study of the program with the $J/\psi$ sample used in the example of this section as well as the $\tau^+\tau^-$, $d\bar{d}$, $u\bar{u}$, $s\bar{s}$, $c\bar{c}$, $B^+B^-$, and $B^0\bar{B}^0$ samples generated at the peak energy of the $\Upsilon(4S)$ resonance.
Each of the used samples consists of one hundred thousand events.
From the left plot in the figure, for all the samples, the number of elapsed seconds grows linearly with the number of processed entries.
This linear pattern is a nice feature.
It guarantees the program has a high rate even in the case of processing huge samples.
For example, the program can process one hundred thousand $J/\psi$ events within five seconds.
Here, we note that the linear pattern is the result of fast searches with unordered maps~\cite{Unordered maps}, as we discuss in Section~\ref{subsection: Algorithm of the program}.
On the other hand, the processing rate of the program varies with the processed samples.
The right plot in Fig.~\ref{figure: Performance study of the program} shows the relationship between the total number of elapsed seconds over the whole sample and the average number of generated particles in an event.
Clearly, a linear pattern is also observed in the plot.
To be specific, with the average number of generated particles in an event increasing by one, the total number of elapsed seconds over the whole sample increases by about 0.56.

%% file: Text/Basics_of_the_program/Output_of_the_program/Output_of_the_program.tex
The program gains the topology information from input data and saves it to output files.
As mentioned in Section \ref{section: Introduction}, the information includes the types of physics processes and the number of processes in each type, involved both in entire samples and in individual events.
We refer to the information at the sample level as topology maps.
In the topology maps, we assign an integer to each type of physics processes as its index.
We term the indices of processes as well as the numbers of processes involved in each type in the individual events as topology tags.

The program outputs topology maps to three different files: one plain text file, one tex source file, and one pdf file, with the same name specified in the card file.
For instance, the three files are ``jpsi\_ta.txt'', ``jpsi\_ta.tex'', and ``jpsi\_ta.pdf'' in the example.
Although in different formats, the three files have the same information.
The pdf file is the easiest to read.
It is converted from the tex source file with the command pdflatex.
The tex source file is convenient to us if we want to change the style of the pdf file to our taste and when we need to copy and paste (parts of) the topology maps to our slides, papers, and so on.
For example, all of the tables displaying topology maps in this paper are taken from associated tex source files.
The plain text file has its own advantage, because the topology maps in it can be checked with text processing commands as well as text editors, and can be used on some occasions as input to the functionality items (see Sections \ref{section: Component analysis} and \ref{section: Signal identification} for details) of another card file.

In addition to the three files for topology maps, one or more root files are output to save topology tags.
The root files only include one TTree object, which is entirely the same as that in the input root files, except for the topology tags inserted in all of its entries.
The number of root files depends on the size of output data. The program switches to one new root file whenever the size of the TTree object in memory exceeds 3 GB.
In the case of the size less than 3 GB, only one root file is output.
While the sole or first root file has the same name as the three files above, more possible root files are denominated with the suffix ``\_n'' (n=1, 2, 3, and so on) appended to the name.
In the example, the first root file is ``jpsi\_ta.root'', and more possible root files would be ``jpsi\_ta\_1.root'', ``jpsi\_ta\_2.root'', ``jpsi\_ta\_3.root'', and so on.

\input{Text/Basics_of_the_program/Output_of_the_program/Decay_trees_and_their_respective_initial-final_states}

In the example of the previous subsection, the program conducts its basic functionality, namely the component analysis over decay trees.
From the 100000 events of the input sample, the program recognizes 17424 decay trees and outputs all of them to the plain text, tex source, and pdf files.
Table \ref{table: Top ten decay trees and their respective final states} only shows the top ten decay trees and their respective final states listed in the output pdf file.
With the help of the symbolic expressions, the components of the sample are clearly displayed in the table, which brings great convenience to us in examining the signals and backgrounds involved in the sample.
In the table, ``rowNo'', ``iDcyTr'', ``nEtr'', and ``nCEtr'' are abbreviations for the row number, index of decay tree, number of entries of decay tree, and number of the cumulative entries from the first to the current decay trees, respectively.
The values of ``iDcyTr'' are assigned from small to large in the program but listed according to the values of ``nEtr'' from large to small in the table. This is the reason why they are not in natural order like the values of ``rowNo''.
Since $J/\psi$ is the only root particle for the $J/\psi$ sample, the production branch $e^+ e^- \rightarrow J/\psi$ is omitted to save page space.
Similar rules also apply to other samples with only one root particle.
Considering $\pi^0$ has a very large production rate and approximatively 99\% of it decays to $\gamma \gamma$, the program is designed to discard the decay $\pi^0 \to \gamma \gamma$ by default at the early phase of processing the input data.
As a result, $\pi^0 \to \gamma \gamma$ does not show itself in the table.
Besides, the superscripts ``$f$'' and ``$F$'' in $\gamma^f$ and $\gamma^F$ indicate the final state radiation effect.
Here, we note that $\gamma^f$ is translated from a special input PDG code ($-$22 in this example), and $\gamma^F$ owns a normal PDG code 22 and meanwhile has at least one $e^{\pm}$, $\mu^{\pm}$, $\pi^{\pm}$, $K^{\pm}$, $p$, or $\bar{p}$ sister.

\begin{figure}[htbp!]
\centering
\includegraphics[width=0.75\textwidth]{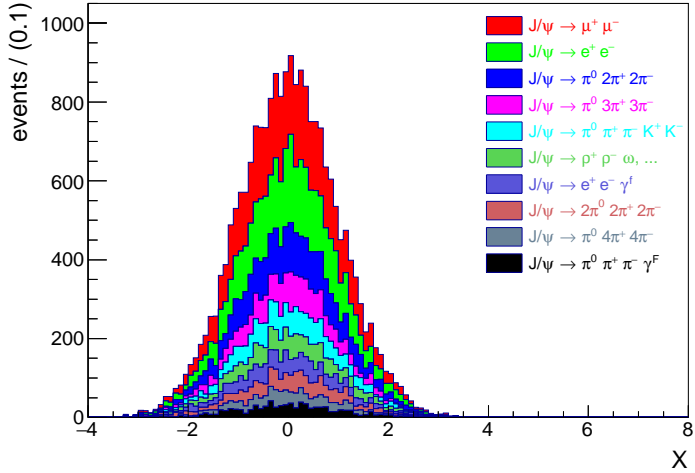}
\caption{Distribution of X accumulated over the top ten decay trees. In the legend entry ``$J/\psi \rightarrow \rho^{+} \rho^{-} \omega$, ...'', the dots ``...'' represent the secondary decay branches: \, $\rho^{+} \rightarrow \pi^{0} \pi^{+}$, \, $\rho^{-} \rightarrow \pi^{0} \pi^{-}$, \,$\omega \rightarrow \pi^{0} \pi^{+} \pi^{-}$.}
\label{figure: Distribution of X accumulated over the top ten decay trees}
\end{figure}

In the table, ``iDcyTr'' is the topology tag for decay trees.
Thus, it is also saved in the TTree objects of the output root file, together with other quantities for physics analysis.
Therefore, it can be used to pick out the entries of specific decay trees and then examine the distributions of the other quantities over the decay trees.
In the example, besides the raw topology truth information, only a random variable following the standardized normal distribution, namely X, is stored in the input root files and thus copied by default to the output root file.
Though not a genuine variable for physics analysis, X is quite good to illustrate the usage of the topology tag.
Figure~\ref{figure: Distribution of X accumulated over the top ten decay trees} shows the distribution of X accumulated over the top ten decay trees.
The figure is drawn with the root script \\
\ \\
\begin{normalsize}
\centerline{examples/in\_the\_paper/ex\_for\_tb\_01/draw\_X/v2/draw\_X.C, }
\end{normalsize}
\ \\
where, for example, a statement equivalent to \\
\ \\
\begin{normalsize}
\centerline{chain--\textgreater Draw(``X \textgreater\textgreater h0'', ``iDcyTr==6'')}
\end{normalsize}
\ \\
is used to import X over the decay tree $J/\psi \rightarrow \mu^{+} \mu^{-}$ from the output root file to the histogram named h0.
With such a figure, we can clearly see the contribution of each decay tree.
Particularly, we can get to know whether a decay tree has a peak contribution or a contribution mainly distributed in a different region.
Based on these distributions, we can get a better understanding of our signals and backgrounds, and thus optimize event selection criteria by applying new requirements on the displayed quantities.

%% file: Text/Basics_of_the_program/Output_of_the_program/Decay_trees_and_their_respective_initial-final_states.tex
\begin{footnotesize}
\begin{centering}

\setcounter{rownumbers}{0}
\begin{longtable}{cllcccc}
\tablecaption{Top ten decay trees and their respective final states. \label{table: Top ten decay trees and their respective final states}}
\tableheader{rowNo & \thead{decay tree} & \thead{decay final state} & \topoTags{iDcyTr & }nEtr & nCEtr \\}

\rn & \makecell[l]{ $
J/\psi \rightarrow \mu^{+} \mu^{-}
$ } & $
\mu^{+} \mu^{-}
$ & \topoTags{6 & }5269 & 5269 \EOL

\rn & \makecell[l]{ $
J/\psi \rightarrow e^{+} e^{-}
$ } & $
e^{+} e^{-}
$ & \topoTags{4 & }4513 & 9782 \EOL

\rn & \makecell[l]{ $
J/\psi \rightarrow \pi^{0} \pi^{+} \pi^{+} \pi^{-} \pi^{-}
$ } & $
\pi^{0} \pi^{+} \pi^{+} \pi^{-} \pi^{-}
$ & \topoTags{0 & }2850 & 12632 \EOL

\rn & \makecell[l]{ $
J/\psi \rightarrow \pi^{0} \pi^{+} \pi^{+} \pi^{+} \pi^{-} \pi^{-} \pi^{-}
$ } & $
\pi^{0} \pi^{+} \pi^{+} \pi^{+} \pi^{-} \pi^{-} \pi^{-}
$ & \topoTags{2 & }1895 & 14527 \EOL

\rn & \makecell[l]{ $
J/\psi \rightarrow \pi^{0} \pi^{+} \pi^{-} K^{+} K^{-}
$ } & $
\pi^{0} \pi^{+} \pi^{-} K^{+} K^{-}
$ & \topoTags{20 & }1698 & 16225 \EOL

\rn & \makecell[l]{ $
J/\psi \rightarrow \rho^{+} \rho^{-} \omega ,
\rho^{+} \rightarrow \pi^{0} \pi^{+} ,
$ \\ $
\ \ \ \ \ \ \rho^{-} \rightarrow \pi^{0} \pi^{-} ,
\omega \rightarrow \pi^{0} \pi^{+} \pi^{-}
$ } & $
\pi^{0} \pi^{0} \pi^{0} \pi^{+} \pi^{+} \pi^{-} \pi^{-}
$ & \topoTags{19 & }1453 & 17678 \EOL

\rn & \makecell[l]{ $
J/\psi \rightarrow e^{+} e^{-} \gamma^{f}
$ } & $
e^{+} e^{-} \gamma^{f}
$ & \topoTags{70 & }1222 & 18900 \EOL

\rn & \makecell[l]{ $
J/\psi \rightarrow \pi^{0} \pi^{0} \pi^{+} \pi^{+} \pi^{-} \pi^{-}
$ } & $
\pi^{0} \pi^{0} \pi^{+} \pi^{+} \pi^{-} \pi^{-}
$ & \topoTags{127 & }1161 & 20061 \EOL

\rn & \makecell[l]{ $
J/\psi \rightarrow \pi^{0} \pi^{+} \pi^{+} \pi^{+} \pi^{+} \pi^{-} \pi^{-} \pi^{-} \pi^{-}
$ } & $
\pi^{0} \pi^{+} \pi^{+} \pi^{+} \pi^{+} \pi^{-} \pi^{-} \pi^{-} \pi^{-}
$ & \topoTags{234 & }836 & 20897 \EOL

\rn & \makecell[l]{ $
J/\psi \rightarrow \pi^{0} \pi^{0} \pi^{+} \pi^{-} \gamma^{F}
$ } & $
\pi^{0} \pi^{0} \pi^{+} \pi^{-} \gamma^{F}
$ & \topoTags{43 & }792 & 21689 \\ \hline

\end{longtable}

\end{centering}
\end{footnotesize}

%% file: Text/Basics_of_the_program/Validation_of_the_program/Validation_of_the_program.tex
The decay trees displayed in Table~\ref{table: Top ten decay trees and their respective final states} are relatively simple, and we can check their correctness by examining the input data directly.
To validate the program generally, we need to do input and output checks, where some arbitrary physics processes are generated as the input of the program.
The output has to be consistent with the input; otherwise, there must be some bugs in the program and we have to fix them.
A large number of such checks have been performed in the development and application of the program, and some of them can be found under the sub-directory ``examples/validation'' of the package.
These checks are divided into two groups: standalone and combined.
In the standalone checks, forty exclusive $J/\psi$ and $\Upsilon(4S)$ decays modeled with the EvtGen~\cite{EvtGen} generator are used to test the functionality of resolving decay trees.
In the combined checks, randomly combined samples of these exclusive decays are used for verifying the functionalities of counting and tagging decay trees.
The output agrees with the input in all the checks, which indicates the correctness of the program.

%% file: Text/Component_analysis/Component_analysis.tex
Component analysis is the primary functionality of the program.
It is developed mainly for the background analysis involved in our physics studies.
We perform it over decay trees in the previous example.
Also, it can be carried out as follows: over decay initial-final states; with specified particles to check their decay branches, cascade decay branches, and decay final states; with specified inclusive decay branches to examine their exclusive components; and with specified intermediate-resonance-allowed (IRA) decay branches to investigate their inner structures.
In this paper, we only introduce the four representative kinds of component analysis, with each in a subsection (for other kinds of component analysis, please see the user guide attached in the package of the program).
For each kind of component analysis, one item is designed and implemented in the program to set related parameters.
In each subsection, we take an example to demonstrate the corresponding setting item and show the resulting topology map.
For easy exposition, all of the essential topology tags involved in the four representative kinds of component analysis functionalities are presented in another separate subsection, namely the last subsection.

Similar to the case over decay trees, to perform the component analysis over decay initial-final states, we only need to input a positive option ``Y'' to the corresponding item.
Different from the former two kinds, to carry out the latter five kinds of component analysis, we have to explicitly specify one or more desired particles, inclusive decay branches, or IRA decay branches in the associated items.
In the following examples, two particles or decay branches are set to illustrate the use of these items, but only the topology map related to one of them is shown to save space in the paper.

In addition to the indispensable parameters, two sorts of common optional parameters can be set in the items.
The first sort is designed for all the seven kinds of component analysis to restrict the maximum number of components output to the plain text, tex source, and pdf files.
Without the optional parameters, all components will be output.
This is fine if the number of components is not massive.
In cases of too many (around ten thousand or more) components, it takes a long time for the program to output the components to the plain text and tex source files as well as to get the pdf file from the tex source file.
In such cases, it also takes up a large disk space to save these components in the output files.
Considering further that the posterior components are generally unimportant and our time and energy to examine them are limited, it is better to set a maximum to the number of output components.
To save space in the paper, we set the maximum number to five in the following examples.

The second sort of optional parameters are developed for the latter five kinds of component analysis to assign meaningful aliases to the specified particles, inclusive decay branches, and IRA decay branches.
By default, the indices 0, 1, 2, and so on are used to tag the particles and decay branches in the names of the TBranch objects appended in the TTree object of the output root files.
This is fine, but it is significative to replace the indices with meaningful aliases, particularly in cases of many specified particles or decay branches.

%% file: Text/Component_analysis/Decay_trees/Decay_trees.tex
Component analysis over decay trees is the basic kind of topology analysis.
It is quite useful to study the backgrounds involved in our research works where the signals are the complete decay trees fully reconstructed from final state particles.
It has already been widely performed in the BESIII experiment, as illustrated in the previous section with the $J/\psi$ example.
This subsection introduces it further with the available optional settings using the $\Upsilon(4S)$ sample.
The following example shows the associated item with the maximum number of output components set to five.
In the item, a third parameter is also filled and set to ``Y''.
With the setting, the decay final states in the output pdf file are put under their respective decay trees, rather than in a column next to that for decay trees.
It is recommended to use this optional parameter in cases there are too many (about ten or more) particles in some final states.
Here, we note that the symbol ``$-$'' can be used as a placeholder for the maximum number of output components, if only the third parameter is desired.
\ \\
\begin{footnotesize}

\% Component analysis --- decay trees

\{

\quad Y \quad 5 \quad Y

\}

\end{footnotesize}
\

Component analysis over decay trees is one kind of the most time-consuming topology analysis tasks.
To check further the efficiency of the program, the progress of running this example, in addition to the example in Section \ref{subsection: Execution of the program}, is illustrated in the plots of Fig. \ref{figure: Performance study of the program} as well.
In these plots, the timing data from this example are marked with the legend entry ``$B^0\bar{B}^0$''.
Since the decay of the $\Upsilon(4S)$ resonance is more complex than that of the $J/\psi$ resonance, it takes more than twenty seconds for the program to process one hundred thousand events in this example.
Nonetheless, the program still has a high processing rate.

Table \ref{table: Decay trees and their respective initial-final states} shows the decay trees.
In the table, while the first five decay trees are listed exclusively in the main part, the rest decay trees are only summarized inclusively at the bottom row.
Here, we note that the events are not densely populated over the first five decay trees because the inclusive $\Upsilon(4S)$ sample used here is not selected beforehand with any requirements.
In the symbolic expressions of decay initial-final states, the dashed right arrow ($\dashrightarrow$) instead of the plain right arrow ($\rightarrow$) is used, in order to reflect that the initial states do not necessarily decay to the final states in a direct way.

\input{Text/Component_analysis/Decay_trees/Decay_trees_and_their_respective_initial-final_states}

%% file: Text/Component_analysis/Decay_trees/Decay_trees_and_their_respective_initial-final_states.tex
\begin{footnotesize}
\begin{centering}

\setcounter{rownumbers}{0}
\begin{longtable}{clcccc}
\tablecaption{Decay trees and their respective initial-final states. \label{table: Decay trees and their respective initial-final states}}
\tableheader{rowNo & \thead{decay tree \\ (decay initial-final states)} & \topoTags{iDcyTr & }nEtr & nCEtr \\}

\rn & \makecell[l]{ $
\Upsilon(4S) \rightarrow B^{0} \bar{B}^{0} ,
B^{0} \rightarrow e^{+} \nu_{e} D^{*-} \gamma^{F} ,
\bar{B}^{0} \rightarrow \mu^{-} \bar{\nu}_{\mu} D^{*+} ,
D^{*-} \rightarrow \pi^{-} \bar{D}^{0} ,
$ \\ $
D^{*+} \rightarrow \pi^{+} D^{0} ,
\bar{D}^{0} \rightarrow \pi^{0} \pi^{-} K^{+} ,
D^{0} \rightarrow \pi^{0} \pi^{+} K^{-}
$ \\ ($
\Upsilon(4S) \dashrightarrow e^{+} \nu_{e} \mu^{-} \bar{\nu}_{\mu} \pi^{0} \pi^{0} \pi^{+} \pi^{+} \pi^{-} \pi^{-} K^{+} K^{-} \gamma^{F}
$) } & \topoTags{20870 & }3 & 3 \EOLTWO

\rn & \makecell[l]{ $
\Upsilon(4S) \rightarrow B^{0} \bar{B}^{0} ,
B^{0} \rightarrow \pi^{0} \pi^{+} \pi^{+} \rho^{-} D^{-} ,
\bar{B}^{0} \rightarrow \mu^{-} \bar{\nu}_{\mu} D^{*+} ,
\rho^{-} \rightarrow \pi^{0} \pi^{-} ,
$ \\ $
D^{-} \rightarrow \pi^{-} \pi^{-} K^{+} ,
D^{*+} \rightarrow \pi^{+} D^{0} ,
D^{0} \rightarrow K_{L}^{0} \pi^{+} \pi^{-}
$ \\ ($
\Upsilon(4S) \dashrightarrow \mu^{-} \bar{\nu}_{\mu} \pi^{0} \pi^{0} K_{L}^{0} \pi^{+} \pi^{+} \pi^{+} \pi^{+} \pi^{-} \pi^{-} \pi^{-} \pi^{-} K^{+}
$) } & \topoTags{5295 & }2 & 5 \EOLTWO

\rn & \makecell[l]{ $
\Upsilon(4S) \rightarrow B^{0} \bar{B}^{0} ,
B^{0} \rightarrow \mu^{+} \nu_{\mu} D^{*-} ,
\bar{B}^{0} \rightarrow e^{-} \bar{\nu}_{e} D^{+} ,
D^{*-} \rightarrow \pi^{-} \bar{D}^{0} ,
$ \\ $
D^{+} \rightarrow e^{+} \nu_{e} \bar{K}^{*} ,
\bar{D}^{0} \rightarrow \pi^{0} \pi^{+} \pi^{-} K_{S}^{0} ,
\bar{K}^{*} \rightarrow \pi^{0} \bar{K}^{0} ,
K_{S}^{0} \rightarrow \pi^{+} \pi^{-} ,
\bar{K}^{0} \rightarrow K_{L}^{0}
$ \\ ($
\Upsilon(4S) \dashrightarrow e^{+} e^{-} \nu_{e} \bar{\nu}_{e} \mu^{+} \nu_{\mu} \pi^{0} \pi^{0} K_{L}^{0} \pi^{+} \pi^{+} \pi^{-} \pi^{-} \pi^{-}
$) } & \topoTags{11954 & }2 & 7 \EOLTWO

\rn & \makecell[l]{ $
\Upsilon(4S) \rightarrow B^{0} \bar{B}^{0} ,
B^{0} \rightarrow e^{+} \nu_{e} D^{*-} ,
\bar{B}^{0} \rightarrow \pi^{0} \pi^{-} \omega D^{+} ,
D^{*-} \rightarrow \pi^{-} \bar{D}^{0} ,
$ \\ $
\omega \rightarrow \pi^{0} \pi^{+} \pi^{-} ,
D^{+} \rightarrow e^{+} \nu_{e} \pi^{+} K^{-} ,
\bar{D}^{0} \rightarrow \pi^{0} \pi^{-} K^{+}
$ \\ ($
\Upsilon(4S) \dashrightarrow e^{+} e^{+} \nu_{e} \nu_{e} \pi^{0} \pi^{0} \pi^{0} \pi^{+} \pi^{+} \pi^{-} \pi^{-} \pi^{-} \pi^{-} K^{+} K^{-}
$) } & \topoTags{14345 & }2 & 9 \EOLTWO

\rn & \makecell[l]{ $
\Upsilon(4S) \rightarrow B^{0} \bar{B}^{0} ,
B^{0} \rightarrow \mu^{+} \nu_{\mu} D^{*-} ,
\bar{B}^{0} \rightarrow e^{-} \bar{\nu}_{e} D^{*+} \gamma^{F} ,
D^{*-} \rightarrow \pi^{-} \bar{D}^{0} ,
$ \\ $
D^{*+} \rightarrow \pi^{0} D^{+} ,
\bar{D}^{0} \rightarrow \pi^{-} K^{+} ,
D^{+} \rightarrow e^{+} \nu_{e} \bar{K}^{*} ,
\bar{K}^{*} \rightarrow \pi^{+} K^{-}
$ \\ ($
\Upsilon(4S) \dashrightarrow e^{+} e^{-} \nu_{e} \bar{\nu}_{e} \mu^{+} \nu_{\mu} \pi^{0} \pi^{+} \pi^{-} \pi^{-} K^{+} K^{-} \gamma^{F}
$) } & \topoTags{15332 & }2 & 11 \EOLTWO

rest & \makecell[l]{ $
\Upsilon(4S) \rightarrow \rm{others \  (99980 \  in \  total)}
$ \\ ($
\Upsilon(4S) \dashrightarrow \rm{corresponding\ to\ others}
$) } & \topoTags{--- & }99989 & 100000 \\ \hline

\end{longtable}

\end{centering}
\end{footnotesize}

%% file: Text/Component_analysis/Decay_initial-final_states/Decay_initial-final_states.tex
On some occasions, we need to investigate the decay initial-final states of backgrounds for some sophisticated physics analyses.
Particularly, it is necessary to differentiate the following two fundamental types of backgrounds: the one with the same initial-final states as the signal, and the other with different initial-final states from the signal.
While the latter type of backgrounds needs to be suppressed as much as possible, the former type usually needs to be kept to study more physical effects, for example, the interference effect.
Besides, examining the decay initial-final states of backgrounds sheds light on the misjudgment of final state particles at the level of signal candidates.
Below is an example demonstrating the related item with the maximum number of output components set to five.
\ \\
\begin{footnotesize}

\% Component analysis --- decay initial-final states

\{

\quad Y \quad 5

\}

\end{footnotesize}
\

\noindent The decay initial-final states are displayed in Table \ref{table: Decay initial-final states}.
The layout of the table is similar to that of Table \ref{table: Decay trees and their respective initial-final states}, which shows the decay trees.

\input{Text/Component_analysis/Decay_initial-final_states/Table_of_decay_initial-final_states}

%% file: Text/Component_analysis/Decay_initial-final_states/Table_of_decay_initial-final_states.tex
\begin{footnotesize}
\begin{centering}

\setcounter{rownumbers}{0}
\begin{longtable}{clccc}
\tablecaption{Decay initial-final states. \label{table: Decay initial-final states}}
\tableheader{rowNo & \thead{decay initial-final states} & \topoTags{iDcyIFSts & }nEtr & nCEtr \\}

\rn & $ \Upsilon(4S) \dashrightarrow \mu^{+} \nu_{\mu} \pi^{0} \pi^{0} \pi^{0} \pi^{+} \pi^{+} \pi^{+} \pi^{-} \pi^{-} \pi^{-} \pi^{-} K^{+} K^{-} $ & \topoTags{41 & }18 & 18 \EOL

\rn & $ \Upsilon(4S) \dashrightarrow \pi^{0} \pi^{0} \pi^{0} \pi^{0} \pi^{0} \pi^{+} \pi^{+} \pi^{+} \pi^{+} \pi^{+} \pi^{-} \pi^{-} \pi^{-} \pi^{-} \pi^{-} K^{+} K^{-} $ & \topoTags{887 & }18 & 36 \EOL

\rn & $ \Upsilon(4S) \dashrightarrow \mu^{-} \bar{\nu}_{\mu} \pi^{0} \pi^{0} \pi^{0} \pi^{0} \pi^{+} \pi^{+} \pi^{+} \pi^{+} \pi^{+} \pi^{-} \pi^{-} \pi^{-} \pi^{-} K^{+} K^{-} $ & \topoTags{3350 & }18 & 54 \EOL

\rn & $ \Upsilon(4S) \dashrightarrow \pi^{0} \pi^{0} \pi^{0} \pi^{0} \pi^{0} \pi^{0} \pi^{+} \pi^{+} \pi^{+} \pi^{+} \pi^{+} \pi^{+} \pi^{-} \pi^{-} \pi^{-} \pi^{-} \pi^{-} \pi^{-} K^{+} K^{-} $ & \topoTags{1215 & }17 & 71 \EOL

\rn & $ \Upsilon(4S) \dashrightarrow \pi^{0} \pi^{0} \pi^{0} \pi^{0} \pi^{0} \pi^{0} K_{L}^{0} \pi^{+} \pi^{+} \pi^{+} \pi^{+} \pi^{+} \pi^{-} \pi^{-} \pi^{-} \pi^{-} K^{-} $ & \topoTags{1207 & }17 & 88 \EOL

rest & $ \Upsilon(4S) \dashrightarrow \rm{others \  (78208 \  in \  total)} $ & \topoTags{--- & }99912 & 100000 \\ \hline

\end{longtable}

\end{centering}
\end{footnotesize}

%% file: Text/Component_analysis/Decay_branches_of_particles/Decay_branches_of_particles.tex
The invariant mass constraint is one of the most frequently used event selection requirements in high energy physics experiments.
With the requirement applied to certain particle, the main backgrounds (especially the peaking ones) to its signal decay mode are very likely to be its other decay modes.
In this case, it is significant to examine the decay branches of the particle.
The following example shows the associated item with the two particles $D^{*+}$ and $J/\psi$ set as research objects.
In the item, each row holds the information of a specified particle, and the first, second and third columns are the textual expressions, aliases, and maximum numbers of output components, respectively.
As we introduce at the beginning part of this section, the aliases and maximum numbers of output components are both optional.
Here, we note that the symbol ``$-$'' can also be used as a placeholder for an unassigned alias, if only the maximum number of output components is desired.
\ \\
\begin{footnotesize}

\% Component analysis --- decay branches of particles

\{

\quad D*+ \quad Dsp \quad 5

\quad J/psi \quad Jpsi \quad 5

\}

\end{footnotesize}
\ \\
Table \ref{table: Decay branches of DSP} shows the decay branches of $D^{*+}$.
From the table,  only four decay branches of $D^{*+}$ are found in the input inclusive MC sample.
Since there is likely one or more cases of $D^{*+}$ decays in one input entry, ``nCase'' and ``nCCase'', instead of ``nEtr'' and ``nCEtr'', are used in the table in order to accurately indicate what we are counting are the numbers of $D^{*+}$ decays, rather than the numbers of entries involving the $D^{*+}$ decays.

\input{Text/Component_analysis/Decay_branches_of_particles/Decay_branches_of_DSP}

It is worth mentioning here that, in addition to decay branches, production branches and mothers of specified particles can also be examined with the program.
One can make the program execute the two functionalities by replacing ``decay branches'' in the prompt of the item with ``production branches'' and ``mothers'', respectively.

%% file: Text/Component_analysis/Decay_branches_of_particles/Decay_branches_of_DSP.tex
\begin{footnotesize}
\begin{centering}

\setcounter{rownumbers}{0}
\begin{longtable}{clccc}
\tablecaption{Decay branches of $ D^{*+} $. \label{table: Decay branches of DSP}}
\tableheader{rowNo & \thead{decay branch of $ D^{*+} $} & \topoTags{iDcyBrP & }nCase & nCCase \\}

\rn & $ D^{*+} \rightarrow \pi^{+} D^{0} $ & \topoTags{0 & }31180 & 31180 \EOL

\rn & $ D^{*+} \rightarrow \pi^{0} D^{+} $ & \topoTags{1 & }13978 & 45158 \EOL

\rn & $ D^{*+} \rightarrow D^{+} \gamma $ & \topoTags{2 & }700 & 45858 \EOL

\rn & $ D^{*+} \rightarrow \pi^{+} D^{0} \gamma^{F} $ & \topoTags{3 & }28 & 45886 \\ \hline

\end{longtable}

\end{centering}
\end{footnotesize}

%% file: Text/Component_analysis/Inclusive_decay_branches/Inclusive_decay_branches.tex
In some physics studies, we take inclusive decay branches as signals.
In such cases, it is essential to have a basic knowledge of the exclusive components of these inclusive decay branches.
Below is an example demonstrating the related item by investigating the exclusive components of the two inclusive decay branches $\bar{B}^{0} \rightarrow D^{*+}  + anything$ and $B^{0} \rightarrow K_{S}^{0}  + anything$.
In the item, each row holds the information of an inclusive decay branch, and the first, second, and third columns separated with the symbol ``\&'' are the textual expressions, aliases, and maximum numbers of output components, respectively.
As we introduce at the beginning part of this section, the aliases and maximum numbers of output components are both optional.
Here, we note that the symbol ``$-$'' can be used as a placeholder for an unassigned alias, if only the maximum number of output components is desired.
\ \\
\begin{footnotesize}

\% Component analysis --- inclusive decay branches

\{

\quad B0 \  $--$\textgreater \ \  D*+ \quad \& \quad B2Dsp \quad \& \quad 5

\quad B0 \  $--$\textgreater \ \  K\_S0 \quad \& \quad B2Ks \quad \& \quad 5

\}

\end{footnotesize}
\ \\
The exclusive components of $B^{0} \rightarrow K_{S}^{0}  + anything$ are displayed in Table \ref{table: Exclusive components of B0 to KS plus anything}.
From the table, ten exclusive components of the inclusive decay branch are found in the input sample, and the particles denoted with {\it anything} are mainly the traditional charmonium states.

\input{Text/Component_analysis/Inclusive_decay_branches/Exclusive_components_of_B0_to_KS_plus_anything}

%% file: Text/Component_analysis/Inclusive_decay_branches/Exclusive_components_of_B0_to_KS_plus_anything.tex
\begin{footnotesize}
\begin{centering}

\setcounter{rownumbers}{0}
\begin{longtable}{clccc}
\tablecaption{Exclusive components of $ B^{0} \rightarrow K_{S}^{0} + anything $. \label{table: Exclusive components of B0 to KS plus anything}}
\tableheader{rowNo & \thead{exclusive component of $ B^{0} \rightarrow K_{S}^{0} + anything $} & \topoTags{iDcyBrIncDcyBr & }nCase & nCCase \\}

\rn & $ B^{0} \rightarrow K_{S}^{0} J/\psi $ & \topoTags{0 & }45 & 45 \EOL

\rn & $ B^{0} \rightarrow K_{S}^{0} \eta_{c} $ & \topoTags{1 & }40 & 85 \EOL

\rn & $ B^{0} \rightarrow K_{S}^{0} \psi^{\prime} $ & \topoTags{3 & }33 & 118 \EOL

\rn & $ B^{0} \rightarrow K_{S}^{0} \chi_{c1} $ & \topoTags{2 & }20 & 138 \EOL

\rn & $ B^{0} \rightarrow K_{S}^{0} \chi_{c0} $ & \topoTags{4 & }6 & 144 \EOL

rest & $ B^{0} \rightarrow K_{S}^{0} + \rm{others \  (5 \  in \  total)} $ & \topoTags{--- & }9 & 153 \\ \hline

\end{longtable}

\end{centering}
\end{footnotesize}

%% file: Text/Component_analysis/Essential_topology_tags/Essential_topology_tags.tex
\input{Text/Component_analysis/Essential_topology_tags/Essential_topology_tags_involved_in_each_kind_of_component_analysis}

Table \ref{table: Essential topology tags involved in each kind of component analysis} lists and interprets all of the essential topology tags involved in the four kinds of component analysis functionalities presented in this section.
The topology tag for the component analysis over decay initial-final states is iDcyIFSts.
It has a similar interpretation as iDcyTr and is shown in the third column of Table \ref{table: Decay initial-final states}.
For the latter five kinds of component analysis, there are two sorts of topology tags.
The first sort, such as nPDcyBr\_i, records the number of instances of the i$^{\rm th}$ specified particle or decay branch found in each event.
The second sort, for example, iDcyBrP\_i\_j, keeps the associated index of the j$^{\rm th}$ found instance of the i$^{\rm th}$ specified particle or decay branch.
The indices and the decays they stand for can be found in Tables \ref{table: Decay branches of DSP} and \ref{table: Exclusive components of B0 to KS plus anything}.

In the topology tags, ``i'' in ``\_i'' is the default index of the specified particle or decay branch, and it ranges from 0 (included) to the number of specified particles or decay branches (excluded).
If the alias of the particle or decay branch is also specified, the index ``i'' will be replaced with the alias.
For example, since ``Dsp'' and ``Jpsi'' are set as the aliases of $D^{*+}$ and $J/\psi$ in the component analysis over their decay branches, the specialized topology tags nPDcyBr\_Dsp and nPDcyBr\_Jpsi, instead of the default ones nPDcyBr\_0 and nPDcyBr\_1, are used to store the numbers of $D^{*+}$ and $J/\psi$ found in each event.

Besides, ``j'' in ``\_j'' is the default index of the found instance of certain particle or decay branch in an event, and it ranges from 0 (included) to the sample-level maximum of the number of instances found in each event (excluded).
For example, the maximum of the number of $D^{*+}$ found in each event is two for the whole sample, and thus two topology tags iDcyBrP\_Dsp\_0 and iDcyBrP\_Dsp\_1 are employed to store the indices of $D^{*+}$ decay branches.
These indices range from 0 (included) to the number of the types of $D^{*+}$ decay branches found in the samples (excluded).
In the events with only one $D^{*+}$, iDcyBrP\_Dsp\_1 is assigned with the default value $-1$; in the events that have no $D^{*+}$, the default value $-1$ is assigned to both iDcyBrP\_Dsp\_0 and iDcyBrP\_Dsp\_1.

%% file: Text/Component_analysis/Essential_topology_tags/Essential_topology_tags_involved_in_each_kind_of_component_analysis.tex
\begin{table}[htbp!]
\centering
\footnotesize
\caption{Essential topology tags involved in each kind of component analysis.}
\label{table: Essential topology tags involved in each kind of component analysis}
\begin{tabular}{lll}
\hline
Component type & Topology tag & Interpretation \\
\hline
Decay trees & iDcyTr & index of decay tree \\
Decay initial-final states & iDcyIFSts & index of decay initial-final states \\
Decay branches of particles & nPDcyBr\_i & number of particle$_{\rm i}$s (or its decay branches) \\
& iDcyBrP\_i\_j & index of decay branch of the ${\rm j}^{\rm th}$ particle$_{\rm i}$ \\
Inclusive decay branches & nIncDcyBr\_i & number of inclusive decay branch$_{\rm i}$es \\
& iDcyBrIncDcyBr\_i\_j & index of decay branch of the ${\rm j}^{\rm th}$ inclusive decay branch$_{\rm i}$ \\
\hline
\end{tabular}
\label{table: Essential topology tags involved in each kind of component analysis}
\end{table}

%% file: Text/Signal_identification/Signal_identification.tex
Signal identification is the other functionality of the program.
Though relatively simple, it can help us identify the ``signals'' we desire directly, quickly, and easily.
Here, the ``signals'' are not confined to the authentic signals in our research works but can be any physics processes of interests, particularly some important backgrounds we concern.
At present, the seven basic kinds of signals that can be identified with the program are as follows:
(1) decay trees,
(2) decay initial-final states,
(3) particles,
(4) (regular) decay branches,
(5) cascade decay branches,
(6) inclusive decay branches, and
(7) IRA decay branches.
For each kind of signals, one item is developed to specify related parameters.
In this paper, we only introduce the former four kinds of signal identification, with each in a subsection (for the latter three kinds of signal identification, please see the user guide attached in the package of the program).
In each subsection, we take an example to demonstrate the related setting item and show the obtained topology map.
For easy exposition, all of the essential topology tags involved in the former four kinds of signal identification functionalities are presented in another separate subsection, that is, the last subsection.

Similar to the cases of the latter five kinds of component analysis, one or more signals can be specified in each of the signal identification items, and two signals are set in the following examples to illustrate the use of the items. Besides, meaning aliases can also be optionally assigned to the specified signals so as to better tag them in the names of the TBranch objects appended in the TTree object of the output root files.

%% file: Text/Signal_identification/Decay_trees/Decay_trees.tex
Sometimes, we need to identify certain decay trees. 
The following example shows the associated item with the first two decay trees listed in Table \ref{table: Decay trees and their respective initial-final states} set as signals.
In the item, each row holds a decay branch in the decay trees, and the first, second, and third columns separated with the symbol ``\&'' are the indices, textual expressions, and mother indices of the decay branches, respectively.
The decay branches with index 0 indicate the beginning of new decay trees, and their mother indices are equal to $-1$, suggesting they have no mother branches because they are the first decay branches of the decay trees.
Besides, the name of each decay tree can be optionally filled in the fourth column of its first decay branch.
Similar to the third parameter in the item for the component analysis over decay trees (see Section \ref{subsection: Component analysis --- decay trees}), a ``Y'' can be optionally filled in the fifth column of the first decay branch of the first decay tree, to adjust the positions of decay final states in the output pdf file.
\ \\
\begin{footnotesize}

\% Signal identification --- decay trees

\{

\quad 0 \quad \& \quad Upsilon(4S) \  $--$\textgreater \ \  B0 \  anti-B0 \quad \& \quad $-$1 \quad \& \quad 1stDcyTrInTb2 \quad \& \quad Y

\quad 1 \quad \& \quad B0 \  $--$\textgreater \ \  e+ \  nu\_e \  D*$-$ \  gamma \quad \& \quad 0

\quad 2 \quad \& \quad anti-B0 \  $--$\textgreater \ \  mu$-$ \  anti-nu\_mu \  D*+ \quad \& \quad 0

\quad 3 \quad \& \quad D*$-$ \  $--$\textgreater \ \  pi$-$ \  anti-D0 \quad \& \quad 1

\quad 4 \quad \& \quad D*+ \  $--$\textgreater \ \  pi+ \  D0 \quad \& \quad 2

\quad 5 \quad \& \quad anti-D0 \  $--$\textgreater \ \  pi0 \  pi$-$ \  K+ \quad \& \quad 3

\quad 6 \quad \& \quad D0 \  $--$\textgreater \ \  pi0 \  pi+ \  K$-$ \quad \& \quad 4

\quad

\quad 0 \quad \& \quad Upsilon(4S) \  $--$\textgreater \ \  B0 \  anti-B0 \quad \& \quad $-$1 \quad \& \quad 2ndDcyTrInTb2

\quad 1 \quad \& \quad B0 \  $--$\textgreater \ \  pi0 \  pi+ \  pi$-$ \  rho$-$ \  D$-$ \quad \& \quad 0

\quad 2 \quad \& \quad anti-B0 \  $--$\textgreater \ \  mu$-$ \  anti-nu\_mu \  D*+ \quad \& \quad 0

\quad 3 \quad \& \quad rho$-$ \  $--$\textgreater \ \  pi0 \  pi$-$ \quad \& \quad 1

\quad 4 \quad \& \quad D$-$ \  $--$\textgreater \ \  pi$-$ \  pi$-$ \  K+ \quad \& \quad 1

\quad 5 \quad \& \quad D*+ \  $--$\textgreater \ \  pi+ \  D0 \quad \& \quad 2

\quad 6 \quad \& \quad D0 \  $--$\textgreater \ \  K\_L0 \  pi+ \  pi$-$ \quad \& \quad 5

\}

\end{footnotesize}
\ \\
Table \ref{table: Signal decay trees and their respective initial-final states} shows the resulting topology map.
The results are the same as those displayed in the first two rows of Table \ref{table: Decay trees and their respective initial-final states}.

\input{Text/Signal_identification/Decay_trees/Signal_decay_trees_and_their_respective_initial-final_states}

%% file: Text/Signal_identification/Decay_trees/Signal_decay_trees_and_their_respective_initial-final_states.tex
\begin{scriptsize}
\begin{centering}

\setcounter{rownumbers}{0}
\begin{longtable}{clccc}
\tablecaption{Signal decay trees and their respective initial-final states. \label{table: Signal decay trees and their respective initial-final states}}
\tableheader{rowNo & \thead{signal decay tree \\ (signal decay initial-final states)} & \topoTags{iSigDcyTr & }nEtr & nCEtr \\}

\rn & \makecell[l]{ $
\Upsilon(4S) \rightarrow B^{0} \bar{B}^{0} ,
B^{0} \rightarrow e^{+} \nu_{e} D^{*-} \gamma^{F} ,
\bar{B}^{0} \rightarrow \mu^{-} \bar{\nu}_{\mu} D^{*+} ,
D^{*-} \rightarrow \pi^{-} \bar{D}^{0} ,
$ \\ $
D^{*+} \rightarrow \pi^{+} D^{0} ,
\bar{D}^{0} \rightarrow \pi^{0} \pi^{-} K^{+} ,
D^{0} \rightarrow \pi^{0} \pi^{+} K^{-}
$ \\ ($
\Upsilon(4S) \dashrightarrow e^{+} \nu_{e} \mu^{-} \bar{\nu}_{\mu} \pi^{0} \pi^{0} \pi^{+} \pi^{+} \pi^{-} \pi^{-} K^{+} K^{-} \gamma^{F}
$) } & \topoTags{0 & }3 & 3 \EOLTWO

\rn & \makecell[l]{ $
\Upsilon(4S) \rightarrow B^{0} \bar{B}^{0} ,
B^{0} \rightarrow \pi^{0} \pi^{+} \pi^{+} \rho^{-} D^{-} ,
\bar{B}^{0} \rightarrow \mu^{-} \bar{\nu}_{\mu} D^{*+} ,
\rho^{-} \rightarrow \pi^{0} \pi^{-} ,
$ \\ $
D^{-} \rightarrow \pi^{-} \pi^{-} K^{+} ,
D^{*+} \rightarrow \pi^{+} D^{0} ,
D^{0} \rightarrow K_{L}^{0} \pi^{+} \pi^{-}
$ \\ ($
\Upsilon(4S) \dashrightarrow \mu^{-} \bar{\nu}_{\mu} \pi^{0} \pi^{0} K_{L}^{0} \pi^{+} \pi^{+} \pi^{+} \pi^{+} \pi^{-} \pi^{-} \pi^{-} \pi^{-} K^{+}
$) } & \topoTags{1 & }2 & 5 \\ \hline

\end{longtable}

\end{centering}
\end{scriptsize}

%% file: Text/Signal_identification/Decay_initial-final_states/Decay_initial-final_states.tex
In some cases, we have an interest in some decay initial-final states.
Below is an example demonstrating the related item by taking the first two decay initial-final states listed in Table \ref{table: Decay initial-final states} as signals.
Similar to IRA decay branches, decay initial-final states look like inclusive decay branches.
Hence, except that only two columns are involved in the item, the format of the input to the item for decay initial-final states is identical to that for the component analysis over inclusive decay branches, which is introduced in Section \ref{subsection: Component analysis --- inclusive decay branches}.
As we can see from the example, the numbers of identical particles are supported to be written in front of their textual names in order to simplify the textual expressions of the final states.
\ \\
\begin{footnotesize}

\% Signal identification --- decay initial-final states

\{

\quad Y(4S) \  $--$\textgreater \ \  mu+ \  nu\_mu \  3 \  pi0 \  3 \  pi+ \  4 \  pi$-$ \  K+ \  K$-$  \quad \& \quad 2ndDcyIFStsInTb3

\quad Y(4S) \  $--$\textgreater \ \  5 \  pi0 \  5 \  pi+ \  5 \  pi$-$ \  K+ \  K$-$ \quad \& \quad 2ndDcyIFStsInTb3

\}

\end{footnotesize}
\ \\
The obtained topology map is displayed in Table \ref{table: Signal decay initial-final states}.
The results are identical to those shown in the first two rows of Table \ref{table: Decay initial-final states}.

\input{Text/Signal_identification/Decay_initial-final_states/Signal_decay_initial-final_states}

%% file: Text/Signal_identification/Decay_initial-final_states/Signal_decay_initial-final_states.tex
\begin{scriptsize}
\begin{centering}

\setcounter{rownumbers}{0}
\begin{longtable}{clccc}
\tablecaption{Signal decay initial-final states. \label{table: Signal decay initial-final states}}
\tableheader{rowNo & \thead{signal decay initial-final states} & \topoTags{iSigDcyIFSts2 & }nEtr & nCEtr \\}

\rn & $ \Upsilon(4S) \dashrightarrow \mu^{+} \nu_{\mu} \pi^{0} \pi^{0} \pi^{0} \pi^{+} \pi^{+} \pi^{+} \pi^{-} \pi^{-} \pi^{-} \pi^{-} K^{+} K^{-} $ & \topoTags{0 & }18 & 18 \EOL

\rn & $ \Upsilon(4S) \dashrightarrow \pi^{0} \pi^{0} \pi^{0} \pi^{0} \pi^{0} \pi^{+} \pi^{+} \pi^{+} \pi^{+} \pi^{+} \pi^{-} \pi^{-} \pi^{-} \pi^{-} \pi^{-} K^{+} K^{-} $ & \topoTags{1 & }18 & 36 \\ \hline

\end{longtable}

\end{centering}
\end{scriptsize}

%% file: Text/Signal_identification/Particles/Particles.tex
Occasionally, we may want to identify some particles.
The following example shows the associated item with the two particles $D^{*+}$ and $J/\psi$ set as signals.
Except that only two columns are involved in the item, the format of the input to the item is identical to that for the component analysis over decay branches of particles, which is introduced in Section \ref{subsection: Component analysis --- decay branches of particles}.
\ \\
\begin{footnotesize}

\% Signal identification --- particles

\{

\quad D*+ \quad Dsp

\quad J/psi \quad Jpsi

\}

\end{footnotesize}
\ \\
Table \ref{table: Signal particles} shows the resulting topology map.
As a cross-check, the number of $D^{*+}$s in the table equals that in Table \ref{table: Decay branches of DSP}.

\input{Text/Signal_identification/Particles/Signal_particles}

%% file: Text/Signal_identification/Particles/Signal_particles.tex
\begin{footnotesize}
\begin{centering}

\setcounter{rownumbers}{0}
\begin{longtable}{ccccc}
\tablecaption{Signal particles. \label{table: Signal particles}}
\tableheader{rowNo & \thead{signal particle} & \topoTags{iSigP & }nCase & nCCase \\}

\rn & $ D^{*+} $ & \topoTags{0 & }45886 & 45886 \EOL

\rn & $ J/\psi $ & \topoTags{1 & }2654 & 48540 \\ \hline

\end{longtable}

\end{centering}
\end{footnotesize}

%% file: Text/Signal_identification/Decay_branches/Decay_branches.tex
On some occasions, we have to identify certain regular decay branches.
Below is an example demonstrating the related item by taking the two decay branches $\bar{B}^{0} \rightarrow \mu^{-} \bar{\nu}_{\mu} D^{*+}$ and $B^{0} \rightarrow K_{S}^{0} J/\psi$ as signals.
Since regular decay branches also look like inclusive decay branches, except that only two columns are involved in the item, the format of the input to the item for regular decay branches is identical to that for the component analysis over inclusive decay branches, which is introduced in Section \ref{subsection: Component analysis --- inclusive decay branches}.
\ \\
\begin{footnotesize}

\% Signal identification --- decay branches

\{

\quad B0 \  $--$\textgreater \ \  mu$-$	\  anti-nu\_mu	\  D*+  \quad \& \quad B2munuDsp

\quad B0 \  $--$\textgreater \ \  K\_S0	\  J/psi \quad \& \quad B2KsJpsi

\}

\end{footnotesize}
\ \\
The obtained topology map is displayed Table \ref{table: Signal decay branches}.
For cross-checks, we note that the number of $B^{0} \rightarrow K_{S}^{0} J/\psi$ in the table is equal to that in the first row of Table \ref{table: Exclusive components of B0 to KS plus anything}.

\input{Text/Signal_identification/Decay_branches/Signal_decay_branches}

%% file: Text/Signal_identification/Decay_branches/Signal_decay_branches.tex
\begin{footnotesize}
\begin{centering}

\setcounter{rownumbers}{0}
\begin{longtable}{clccc}
\tablecaption{Signal decay branches. \label{table: Signal decay branches}}
\tableheader{rowNo & \thead{signal decay branch} & \topoTags{iSigDcyBr & }nCase & nCCase \\}

\rn & $ \bar{B}^{0} \rightarrow \mu^{-} \bar{\nu}_{\mu} D^{*+} $ & \topoTags{0 & }4154 & 4154 \EOL

\rn & $ B^{0} \rightarrow K_{S}^{0} J/\psi $ & \topoTags{1 & }45 & 4199 \\ \hline

\end{longtable}

\end{centering}
\end{footnotesize}

%% file: Text/Signal_identification/Essential_topology_tags/Essential_topology_tags.tex
Table \ref{table: Essential topology tags involved in each kind of signal identification} summarizes and explains all of the essential topology tags involved in the four kinds of signal identification functionalities presented in this section.
For signal decay trees and signal decay initial-final states, there are two sorts of topology tags.
The first sort of tags, iSigDcyTr and iSigDcyIFSts, record the default indices of the specified signal decay trees and signal decay initial-final states.
They have similar interpretations as iDcyTr and iDcyIFSts, and are shown in the third columns of Tables \ref{table: Signal decay trees and their respective initial-final states} and \ref{table: Signal decay initial-final states}.
The second sort of tags, nameSigDcyTr and nameSigDcyIFSts, save the specified aliases of the signal decay trees and signal decay initial-final states.
In cases the aliases are not specified, empty strings will be stored.

For the latter five kinds of signal identification, there is only one sort of topology tags, which records the number of instances of certain specified particle or decay branch found in each event.
Similar to the cases in the latter five kinds of component analysis, in the topology tags, ``i'' in ``\_i'' is the default index of the specified particle or decay branch, and it ranges from 0 (included) to the number of specified particles or decay branches (excluded).
If the alias of the particle or decay branch is also specified, the index ``i'' will be replaced with the alias.

\input{Text/Signal_identification/Essential_topology_tags/Essential_topology_tags_involved_in_each_kind_of_signal_identification}

%% file: Text/Signal_identification/Essential_topology_tags/Essential_topology_tags_involved_in_each_kind_of_signal_identification.tex
\begin{table}[htbp!]
\centering
\footnotesize
\caption{Essential topology tags involved in each kind of signal identification.}
\label{table: Essential topology tags involved in each kind of signal identification}
\begin{tabular}{lll}
\hline
Signal type & Topology tag & Interpretation \\
\hline
Decay trees & iSigDcyTr & index of signal decay tree \\
& nameSigDcyTr & name of signal decay tree \\
Decay initial-final states & iSigDcyIFSts & index of signal decay initial-final states \\
& nameSigDcyIFSts & name of signal decay initial-final states \\
Particles & nSigP\_i & number of signal particle$_{\rm i}$s \\
Decay branches & nSigDcyBr\_i & number of signal decay branch$_{\rm i}$es \\
\hline
\end{tabular}
\label{table: Essential topology tags involved in each kind of signal identification}
\end{table}

%% file: Text/Common_settings/Common_settings.tex
From Sections \ref{section: Component analysis} and \ref{section: Signal identification}, the optional parameters of the functionality items give us more choices and thus help us do our jobs quicker and better.
In addition to these parameters, many optional items are designed and implemented to control the execution of the program in order to meet practical needs.
Unlike the optional parameters, which only affect the individual functionalities to which they belong, the optional items have an impact on all of the functionalities, or at least most of the functionalities.
The current version of the program contains two dozen common setting items on its input, functionalities, and output.
In this paper, we only introduce a part of the items that are crucial to our physics studies (for other items, please see the user guide attached in the package of the program).

%% file: Text/Common_settings/Settings_on_input_entries/Settings_on_input_entries.tex
The program normally processes all of the entries in the input samples, but sometimes only a part of the entries are needed to be (first) processed.
Running the program over a big sample usually takes a long time.
In such a case, it is a good habit to run the program first over a small part of the sample to check possible exceptions, and then over the whole sample if no exceptions are found or after the found exceptions are handled.
Besides, a small number of entries is usually sufficient to do tests in the development of the program.
For these reasons, an item is developed to set up the maximum number of entries to be processed.
Below is an example showing the item with the maximum number set at two thousand.
\ \\
\begin{footnotesize}

\% Maximum number of entries to be processed

\{

\quad 2000

\}

\end{footnotesize}
\

On some occasions, especially in the course of optimizing selection criteria, we need to run the program only over entries satisfying certain requirements. For this purpose, an item is developed to select entries. The following example shows the item with X set in the range ($-$1, 1).
\ \\
\begin{footnotesize}

\% Cut to select entries

\{

\quad (X \textgreater $-$1) \&\& (X \textless 1)

\}

\end{footnotesize}
\ \\
\noindent Notably, only a single-line selection requirement is supported in the item, like the cases in the methods Draw()~\cite{ROOT2} and GetEntries()~\cite{ROOT2} of the class TTree. In spite of this, such a requirement is able to express any requirement with the help of the parentheses ``()'' as well as the logical symbols ``\&\&'', ``$||$'', and ``!''.

%% file: Text/Common_settings/Setting_on_input_decay_branches/Setting_on_input_decay_branches.tex
Normally, the program deals with all of the decay branches in every decay tree.
However, examining all the branches is not always required in practice.
Sometimes, we only concern the first $n$ hierarchies of the branches.
Here, the hierarchy reflects the rank of a decay branch in a decay tree.
For example, in the decay tree
$\Upsilon(4S) \rightarrow B^{0} \bar{B}^{0}$,
$B^{0} \rightarrow e^{+} \nu_{e} D^{*-} \gamma^{F}$,
$\bar{B}^{0} \rightarrow \mu^{-} \bar{\nu}_{\mu} D^{*+}$,
$D^{*-} \rightarrow \pi^{-} \bar{D}^{0}$,
$D^{*+} \rightarrow \pi^{+} D^{0}$,
$\bar{D}^{0} \rightarrow \pi^{0} \pi^{-} K^{+}$,
$D^{0} \rightarrow \pi^{0} \pi^{+} K^{-}$,
the hierarchies of the seven individual branches are 1, 2, 2, 3, 3, 4, and 4, respectively.
The program provides an item to set the maximum hierarchy. Below is an example showing the item with the maximum hierarchy set at one.
\ \\
\begin{footnotesize}

\% Maximum hierarchy of heading decay branches to be processed in each event

\{

\quad 1

\}

\end{footnotesize}
\ \\

\noindent With the setting, the decay branches with hierarchy larger than one will be ignored by the program.
For the component analysis over the decay trees of the $\Upsilon(4S)$ sample, only the first hierarchy of $\Upsilon(4S)$ decay branches are analyzed, and the result is shown in Table \ref{table: Decay trees and their respective initial-final states I}. From the table, not only $\Upsilon(4S) \rightarrow B^{0} \bar{B}^{0}$ but also $\Upsilon(4S) \rightarrow B^{0} B^{0}$ and $\Upsilon(4S) \rightarrow \bar{B}^{0} \bar{B}^{0}$ are seen because of $B^{0}$-$\bar{B}^{0}$ mixing.

\input{Text/Common_settings/Setting_on_input_decay_branches/Decay_trees_and_their_respective_initial-final_states_I}

\noindent Similarly, in the case of the maximum hierarchy set at two, we could get the result of the component analysis over the first two hierarchies of $\Upsilon(4S)$ decay branches, as displayed in Table \ref{table: Decay trees and their respective initial-final states II}.

\input{Text/Common_settings/Setting_on_input_decay_branches/Decay_trees_and_their_respective_initial-final_states_II}

%% file: Text/Common_settings/Setting_on_input_decay_branches/Decay_trees_and_their_respective_initial-final_states_I.tex
\begin{footnotesize}
\begin{centering}

\setcounter{rownumbers}{0}
\begin{longtable}{clcccc}
\tablecaption{Decay trees and their respective initial-final states. \label{table: Decay trees and their respective initial-final states I}}
\tableheader{rowNo & \thead{decay tree \\ (decay initial-final states)} & \topoTags{iDcyTr & }nEtr & nCEtr \\}

\rn & \makecell[l]{ $
\Upsilon(4S) \rightarrow B^{0} \bar{B}^{0}
$ \\ ($
\Upsilon(4S) \dashrightarrow B^{0} \bar{B}^{0}
$) } & \topoTags{0 & }81057 & 81057 \EOL

\rn & \makecell[l]{ $
\Upsilon(4S) \rightarrow B^{0} B^{0}
$ \\ ($
\Upsilon(4S) \dashrightarrow B^{0} B^{0}
$) } & \topoTags{1 & }9487 & 90544 \EOL

\rn & \makecell[l]{ $
\Upsilon(4S) \rightarrow \bar{B}^{0} \bar{B}^{0}
$ \\ ($
\Upsilon(4S) \dashrightarrow \bar{B}^{0} \bar{B}^{0}
$) } & \topoTags{2 & }9456 & 100000 \\ \hline

\end{longtable}

\end{centering}
\end{footnotesize}

%% file: Text/Common_settings/Setting_on_input_decay_branches/Decay_trees_and_their_respective_initial-final_states_II.tex
\begin{footnotesize}
\begin{centering}

\setcounter{rownumbers}{0}
\begin{longtable}{clcccc}
\tablecaption{Decay trees and their respective initial-final states. \label{table: Decay trees and their respective initial-final states II}}
\tableheader{rowNo & \thead{decay tree \\ (decay initial-final states)} & \topoTags{iDcyTr & }nEtr & nCEtr \\}

\rn & \makecell[l]{ $
\Upsilon(4S) \rightarrow B^{0} \bar{B}^{0} ,
B^{0} \rightarrow \mu^{+} \nu_{\mu} D^{*-} ,
\bar{B}^{0} \rightarrow \mu^{-} \bar{\nu}_{\mu} D^{*+}
$ \\ ($
\Upsilon(4S) \dashrightarrow \mu^{+} \mu^{-} \nu_{\mu} \bar{\nu}_{\mu} D^{*+} D^{*-}
$) } & \topoTags{936 & }136 & 136 \EOL

\rn & \makecell[l]{ $
\Upsilon(4S) \rightarrow B^{0} \bar{B}^{0} ,
B^{0} \rightarrow e^{+} \nu_{e} D^{*-} ,
\bar{B}^{0} \rightarrow \mu^{-} \bar{\nu}_{\mu} D^{*+}
$ \\ ($
\Upsilon(4S) \dashrightarrow e^{+} \nu_{e} \mu^{-} \bar{\nu}_{\mu} D^{*+} D^{*-}
$) } & \topoTags{1188 & }112 & 248 \EOL

\rn & \makecell[l]{ $
\Upsilon(4S) \rightarrow B^{0} \bar{B}^{0} ,
B^{0} \rightarrow \mu^{+} \nu_{\mu} D^{*-} ,
\bar{B}^{0} \rightarrow e^{-} \bar{\nu}_{e} D^{*+}
$ \\ ($
\Upsilon(4S) \dashrightarrow e^{-} \bar{\nu}_{e} \mu^{+} \nu_{\mu} D^{*+} D^{*-}
$) } & \topoTags{268 & }110 & 358 \EOL

\rn & \makecell[l]{ $
\Upsilon(4S) \rightarrow B^{0} \bar{B}^{0} ,
B^{0} \rightarrow D^{*-} D_{s}^{*+} ,
\bar{B}^{0} \rightarrow \mu^{-} \bar{\nu}_{\mu} D^{*+}
$ \\ ($
\Upsilon(4S) \dashrightarrow \mu^{-} \bar{\nu}_{\mu} D^{*+} D^{*-} D_{s}^{*+}
$) } & \topoTags{2063 & }72 & 430 \EOL

\rn & \makecell[l]{ $
\Upsilon(4S) \rightarrow B^{0} \bar{B}^{0} ,
B^{0} \rightarrow e^{+} \nu_{e} D^{*-} ,
\bar{B}^{0} \rightarrow e^{-} \bar{\nu}_{e} D^{*+}
$ \\ ($
\Upsilon(4S) \dashrightarrow e^{+} e^{-} \nu_{e} \bar{\nu}_{e} D^{*+} D^{*-}
$) } & \topoTags{95 & }71 & 501 \EOL

rest & \makecell[l]{ $
\Upsilon(4S) \rightarrow \rm{others \  (81609 \  in \  total)}
$ \\ ($
\Upsilon(4S) \dashrightarrow \rm{corresponding\ to\ others}
$) } & \topoTags{--- & }99499 & 100000 \\ \hline

\end{longtable}

\end{centering}
\end{footnotesize}

%% file: Text/Common_settings/Setting_on_charge_conjugation/Setting_on_charge_conjugation.tex
Charge conjugation is an important concept in high energy physics.
By default, charge conjugate objects (particles and decays) are processed separately in the program.
However, we need to handle them together in many physics studies because of the sameness between them.
One can have the program process them together with the item below set to ``Y''.
\ \\
\begin{footnotesize}

\% Process charge conjugate objects together (Two options: Y and N. Default: N)

\{

\quad Y

\}

\end{footnotesize}
\ \\
Performing topology analysis with this setting inserts new topology tags in the output root files and adds new counters to topology maps in the output plain text, tex source, and pdf files.
Tables \ref{table: Topology tags related to charge conjugation involved in each kind of component analysis} and \ref{table: Topology tags related to charge conjugation involved in each kind of signal identification} list and interpret all of the topology tags related to charge conjugation involved in the component analysis and signal identification functionalities, respectively.

As an example, we carry out the component analysis over the decay branches of $D^{*+}$ and $J/\psi$ with the charge conjugation setting.
The resulting topology map of $D^{*+}$ is displayed in Table \ref{table: Decay branches of DSP with cc}.
Besides the columns in Table \ref{table: Decay branches of DSP}, two additional columns with the headers ``nCcCase'' and ``nAllCase'' are inserted in the table.
Here, ``nCcCase'' represents the number of cases involving the charge conjugate particle ($D^{*-}$ in this table), and ``nAllCase'' is the sum of ``nCase'' and ``nCcCase''.

\input{Text/Common_settings/Setting_on_charge_conjugation/Topology_tags_related_to_charge_conjugation_involved_in_each_kind_of_component_analysis}

In the example, besides the essential topology tags ``nPDcyBr\_i'' and ``iDcyBrP\_i\_j'', the following three groups of topology tags related to charge conjugation are also inserted in the output root files: (1) ``iCcPDcyBr\_i'' for all specified particles; (2) ``iCcDcyBrP\_i\_j'' only for self-charge-conjugate particles, such as $J/\psi$; (3) ``nCcPDcyBr\_i'', ``iDcyBrCcP\_i\_j'', and ``nAllPDcyBr\_i'' only for non-self-charge-conjugate particles, such as $D^{*+}$.
Here, ``iCcPDcyBr\_i'' tags whether the i$^{\rm th}$ particle is self-charge-conjugate.
For self-charge-conjugate particles, it has the value 0; for non-self-charge-conjugate particles, it has the value 1.

The topology tag ``iCcDcyBrP\_i\_j'' records the charge conjugation property of the decay branch of the ${\rm j}^{\rm th}$ instance of the ${\rm i}^{\rm th}$ particle.
For self-charge-conjugate decay branches, it has the value 0; for non-self-charge-conjugate decay branches, it has the value 1 or $-$1: while 1 tags the decay branches listed in the topology maps, $-$1 indicates their charge conjugate decay branches.
Whereas the equal values of ``iDcyBrP\_i\_j'' for each decay branch and its charge conjugate decay branch indicate their sameness, the opposite values of ``iCcDcyBrP\_i\_j'' for them reflect their difference.

The topology tag ``iDcyBrCcP\_i\_j'' is designed for the charge conjugate particle of the i$^{\rm th}$ particle (for $D^{*-}$ in this example).
It has a similar meaning as ``iDcyBrP\_i\_j''.
Particularly, the values of ``iDcyBrP\_i\_j'' and ``iDcyBrCcP\_i\_j'' tagging charge conjugate decay branches are equal to each other.
The topology tag ``nCcPDcyBr\_i'' stands for the number of instances (or decay branches) of the charge conjugate particle of the i$^{\rm th}$ particle found in each event, and ``nAllPDcyBr\_i'' is the sum of ``nPDcyBr\_i'' and ``nCcPDcyBr\_i''.

\input{Text/Common_settings/Setting_on_charge_conjugation/Topology_tags_related_to_charge_conjugation_involved_in_each_kind_of_signal_identification}

\input{Text/Common_settings/Setting_on_charge_conjugation/Decay_branches_of_DSP_with_cc}

%% file: Text/Common_settings/Setting_on_charge_conjugation/Topology_tags_related_to_charge_conjugation_involved_in_each_kind_of_component_analysis.tex
\begin{table}[htbp!]
\centering
\footnotesize
\caption{Topology tags related to charge conjugation involved in each kind of component analysis. For the latter five kinds of component analysis, the topology tags in the (1) and (2) groups are only designed for the self-charge-conjugate and non-self-charge-conjugate particles and decay branches, respectively. The acronyms ``cc'' and index$_{\rm cc}$ are short for ``charge conjugate'' and ``charge conjugate index'', respectively.
For self-charge-conjugate objects (particles or decays), the charge conjugate indices have the value 0; for non-self-charge-conjugate objects, they have the value 1 or $-$1: while 1 tags the objects presented in the topology maps, $-$1 indicates their charge conjugate objects.}
\label{table: Topology tags related to charge conjugation involved in each kind of component analysis}
\begin{tabular}{lll}
\hline
Component type & Topology tag & Interpretation \\
\hline
Decay trees & iCcDcyTr & index$_{\rm cc}$ of decay tree \\
\hline
Decay initial-final states & iCcDcyIFSts & index$_{\rm cc}$ of decay initial-final states \\
\hline
& iCcPDcyBr\_i & index$_{\rm cc}$ of particle$_{\rm i}$ \\
& (1) iCcDcyBrP\_i\_j & index$_{\rm cc}$ of decay branch of the ${\rm j}^{\rm th}$ particle$_{\rm i}$ \\
Decay branches & (2) nCcPDcyBr\_i & number of cc particle$_{\rm i}$s (decay branches) \\
of particles & (2) iDcyBrCcP\_i\_j & index of decay branch of the ${\rm j}^{\rm th}$ cc particle$_{\rm i}$ \\
& (2) nAllPDcyBr\_i & number of all particle$_{\rm i}$s (decay branches) \\
\hline
\multirow{5}{*}{Inclusive decay branches} & iCcIncDcyBr\_i & index$_{\rm cc}$ of inclusive decay branch$_{\rm i}$ \\
& (1) iCcDcyBrIncDcyBr\_i\_j & index$_{\rm cc}$ of decay branch of the ${\rm j}^{\rm th}$ inclusive decay branch$_{\rm i}$ \\
& (2) nCcIncDcyBr\_i & number of cc inclusive decay branch$_{\rm i}$es \\
& (2) iDcyBrCcIncDcyBr\_i\_j & index of decay branch of the ${\rm j}^{\rm th}$ cc inclusive decay branch$_{\rm i}$ \\
& (2) nAllIncDcyBr\_i & number of all inclusive decay branch$_{\rm i}$es \\
\hline
\end{tabular}
\label{table: Topology tags related to charge conjugation involved in each kind of component analysis}
\end{table}

%% file: Text/Common_settings/Setting_on_charge_conjugation/Topology_tags_related_to_charge_conjugation_involved_in_each_kind_of_signal_identification.tex
\begin{table}[htbp!]
\centering
\footnotesize
\caption{Topology tags related to charge conjugation involved in each kind of signal identification. For the latter five kinds of signal identification, the topology tags in the (*) groups are only designed for the non-self-charge-conjugate particles and decay branches. The acronyms ``cc'' and index$_{\rm cc}$ are short for ``charge conjugate'' and ``charge conjugate index'', respectively. For self-charge-conjugate objects (particles or decays), the charge conjugate indices have the value 0; for non-self-charge-conjugate objects, they have the value 1 or $-$1: while 1 tags the objects presented in the topology maps, $-$1 indicates their charge conjugate objects.}
\label{table: Topology tags related to charge conjugation involved in each kind of signal identification}
\begin{tabular}{lll}
\hline
Signal type & Topology tag & Interpretation \\
\hline
Decay trees & iCcSigDcyTr & index$_{\rm cc}$ of signal decay tree \\
\hline
Decay initial-final states & iCcSigDcyIFSts & index$_{\rm cc}$ of signal decay initial-final states \\
\hline
\multirow{3}{*}{Particles} & iCcSigP\_i & index$_{\rm cc}$ of signal particle$_{\rm i}$ \\
& (*) nCcSigP\_i & number of cc signal particle$_{\rm i}$s \\
& (*) nAllSigP\_i & number of all signal particle$_{\rm i}$s \\
\hline
\multirow{3}{*}{Decay branches} & iCcSigDcyBr\_i & index$_{\rm cc}$ of signal decay branch$_{\rm i}$ \\
& (*) nCcSigDcyBr\_i & number of cc signal decay branch$_{\rm i}$es \\
& (*) nAllSigDcyBr\_i & number of all signal decay branch$_{\rm i}$es \\
\hline
\end{tabular}
\label{table: Topology tags related to charge conjugation involved in each kind of signal identification}
\end{table}

%% file: Text/Common_settings/Setting_on_charge_conjugation/Decay_branches_of_DSP_with_cc.tex
\begin{footnotesize}
\begin{centering}

\setcounter{rownumbers}{0}
\begin{longtable}{clccccc}
\tablecaption{Decay branches of $ D^{*+} $ (with the charge conjugation setting). \label{table: Decay branches of DSP with cc}}

\tableheader{rowNo & \thead{decay branch of $ D^{*+} $} & \topoTags{iDcyBrP & }nCase & nCcCase & nAllCase & nCCase \\}

\rn & $ D^{*+} \rightarrow \pi^{+} D^{0} $ & \topoTags{0 & }31180 & 31291 & 62471 & 62471 \EOL

\rn & $ D^{*+} \rightarrow \pi^{0} D^{+} $ & \topoTags{1 & }13978 & 14166 & 28144 & 90615 \EOL

\rn & $ D^{*+} \rightarrow D^{+} \gamma $ & \topoTags{2 & }700 & 721 & 1421 & 92036 \EOL

\rn & $ D^{*+} \rightarrow \pi^{+} D^{0} \gamma^{F} $ & \topoTags{3 & }28 & 36 & 64 & 92100 \EOL

\rn & $ D^{*+} \rightarrow \pi^{0} D^{+} \gamma $ & \topoTags{4 & }0 & 1 & 1 & 92101 \\ \hline

\end{longtable}

\end{centering}
\end{footnotesize}

%% file: Text/Common_settings/Setting_on_initial_state_particles/Setting_on_initial_state_particles.tex
In all of the previous examples, the program is applied to the inclusive MC samples in $e^+e^-$ colliding experiments.
Besides, the program can also be used in other types of high energy experiments, for example, the PANDA experiment~\cite{PANDA}, a $p \bar{p}$ annihilation experiment under construction at Darmstadt, Germany.
On these occasions, we have to specify the right initial state particles with the following item to obtain the proper topology maps.
\ \\
\begin{footnotesize}

\% Initial state particles (Default: e$-$ e+)

\{

\quad anti-p$-$ \quad p+

\}

\end{footnotesize}
\

\noindent With the setting, the default initial state $e^+e^-$ is replaced by $p \bar{p}$, as shown in Table~\ref{table: Decay trees and their respective final states (ppbar)}, which displays the results of a component analysis over decay trees of a small $p \bar{p}$ annihilation sample.

\input{Text/Common_settings/Setting_on_initial_state_particles/Decay_trees_and_their_respective_final_states_ppbar}

%% file: Text/Common_settings/Setting_on_initial_state_particles/Decay_trees_and_their_respective_final_states_ppbar.tex
\begin{footnotesize}
\begin{centering}

\setcounter{rownumbers}{0}
\begin{longtable}{cllcccc}
\tablecaption{Decay trees and their respective final states ($p \bar{p}$ annihilation). \label{table: Decay trees and their respective final states (ppbar)}}
\tableheader{rowNo & \thead{decay tree} & \thead{decay final state} & \topoTags{iDcyTr & }nEtr & nCEtr \\}

\rn & \makecell[l]{ $
p \bar{p} \rightarrow p \bar{p}
$ } & $
p \bar{p}
$ & \topoTags{1 & }232 & 232 \EOL

\rn & \makecell[l]{ $
p \bar{p} \rightarrow \pi^{+} \pi^{-} p \bar{p}
$ } & $
\pi^{+} \pi^{-} p \bar{p}
$ & \topoTags{24 & }53 & 285 \EOL

\rn & \makecell[l]{ $
p \bar{p} \rightarrow \pi^{0} p \bar{p}
$ } & $
\pi^{0} p \bar{p}
$ & \topoTags{5 & }35 & 320 \EOL

\rn & \makecell[l]{ $
p \bar{p} \rightarrow \pi^{0} \pi^{+} \pi^{-} p \bar{p}
$ } & $
\pi^{0} \pi^{+} \pi^{-} p \bar{p}
$ & \topoTags{0 & }33 & 353 \EOL

\rn & \makecell[l]{ $
p \bar{p} \rightarrow \pi^{0} \pi^{0} \pi^{0} \pi^{+} \pi^{+} \pi^{-} \pi^{-}
$ } & $
\pi^{0} \pi^{0} \pi^{0} \pi^{+} \pi^{+} \pi^{-} \pi^{-}
$ & \topoTags{39 & }31 & 384 \EOL

rest & \makecell[l]{ $
p \bar{p} \rightarrow \rm{others \  (184 \  in \  total)}
$ } & $
\rm{corresponding\ to\ others}
$ & \topoTags{--- & }616 & 1000 \\ \hline

\end{longtable}

\end{centering}
\end{footnotesize}

%% file: Text/Summary/Summary.tex
We develop a program, namely TopoAna, with C++, ROOT, and LaTeX for the event type analysis of inclusive MC samples in high energy physics experiments.
The program has rich functionalities and aims to solve all kinds of event type analysis tasks.
Meanwhile, it is easy to use and has a high processing rate.
These features make the program a powerful tool to analyze the backgrounds involved in our research works and to identify the physics processes of interests from the inclusive MC samples.

Since it does not rely on any specific software frameworks, the program applies to many high energy physics experiments.
Up to now, it has been put into use in three experiments at $e^+e^-$ colliders: the BESIII, Belle, and Belle~II experiments.
Besides these experiments, it can also be used in other types of experiments, such as the PANDA experiment, a $p \bar{p}$ annihilation experiment.
Also, the program is applicable to the future $e^+e^-$ colliding experiments under research and development, such as the circular electron-positron collider (CEPC)~\cite{CEPC1,CEPC2} experiment in China, the super Charm-$\tau$ factory (SCTF) experiment~\cite{SCTF} in Russia, and the super $\tau$-Charm factory (STCF) experiment~\cite{STCF} in China.
These experiments offer wide space for the application of the program.

On the other hand, we note that the application of the program to some other experiments is limited.
For example, thousands of particles can be produced from dozens of $pp$ collisions in an event of the ATLAS~\cite{ATLAS} and CMS~\cite{CMS} experiments at the LHC~\cite{LHC}; in such cases, there is little point in performing the event type analysis of corresponding MC samples.
Nonetheless, the application scope of the program is still broad.
In particular, it applies to the $e^+e^-$ colliding experiments where at most tens of particles are produced from the annihilation of a pair of $e^+e^-$ in an event.
With more user needs coming out in the future, we will further extend and perfect it to make it more powerful and well-rounded.

%% file: Acknowledgements/Acknowledgements.tex
This work was supported by the National Natural Science Foundation of China [grant numbers 11575017, 11661141008, 11761141009, 11875262, 11975076] and the CAS Center for Excellence in Particle Physics (CCEPP).
In addition, we would like to thank all of the people who have helped us in the development of the program.
We first thank Prof. Changzheng Yuan, Bo Xin, and Haixuan Chen for their help at the early stage of developing the program.
We are particularly grateful to Prof. Xingtao Huang for his comments on the principles and styles of the program, to Remco de Boer for his suggestions on the tex output and the use of GitHub, and to Xi Chen for his discussions on the core algorithms.
We are especially indebted to Prof. Xiqing Hao, Longke Li, Xiaoping Qin, Ilya Komarov, Yubo Li, Guanda Gong, Suxian Li, Junhao Yin, Prof. Xiaolong Wang, and Yeqi Chen for their advice in extending and perfecting the program.
Also, we thank Xi'an Xiong, Runqiu Ma, Wencheng Yan, Sen Jia, Lu Cao, Dong Liu, Hongpeng Wang, Jiawei Zhang, Hongrong Qi, Jiajun Liu, Maoqiang Jing, Yi Zhang, Wei Shan, and Yadi Wang for their efforts in helping us test the program.

%% file: References/References.tex
\bibliographystyle{elsarticle-num}